\documentclass[useAMS,usenatbib]{mn2e}
\usepackage{amsmath}
\usepackage{graphicx}
\usepackage{epstopdf}
\usepackage{aas_macros}
\usepackage{amssymb}

\newcommand{\Msun}{M$_{\odot}$ }

\newcommand{\bvri}{\hbox{$BV\!RI$} }
\newcommand{\nic}{\hbox{$^{56}$Ni }}
\newcommand{\nico}{\hbox{$^{56}$Ni}}
\newcommand{\dmb}{\hbox{$\Delta m_{15}(B)$}}
\newcommand{\dmv}{\hbox{$\Delta m_{15}(V)$}}
\newcommand{\sr}{\hbox{$s_{r}$}}
\newcommand{\sfl}{\hbox{$s_{f}$}}
\newcommand{\noo}{\hbox{61}}
\newcommand{\ion}[2]{#1$\;${\small{#2}}\relax}

\title[Rise Times of Nearby SNe~Ia]{The Rise-Time Distribution of Nearby Type Ia Supernovae}

\author[Ganeshalingam, Li, \& Filippenko]{Mohan Ganeshalingam,\thanks{E-mail: mganesh@astro.berkeley.edu} Weidong
  Li, and Alexei V. Filippenko \\
Department of Astronomy, University of California,
Berkeley, CA 94720-3411, USA\\
}

\begin{document}

\date{Accepted 2011 June 8. Received   2011 June 6; in original form  2010 October 12}

\pagerange{\pageref{firstpage}--\pageref{lastpage}} \pubyear{2011}

\maketitle

\label{firstpage}

\begin{abstract}
We present an analysis of the $B$-band and $V$-band rise-time distributions of nearby Type Ia supernovae (SNe~Ia). Drawing mostly from the recently published Lick Observatory Supernova Search sample of SNe~Ia \citep{ganeshalingam10}, together with other published nearby SNe~Ia with data starting at least one week before maximum light, we use a two-stretch template-fitting method to measure the rise and decline of $BV$ light curves.  Our analysis of \noo\ SNe with high-quality light curves indicates that the longer the time between explosion and maximum light (i.e., the rise time),  the slower the decline of the light curve after maximum. However, SNe with slower post-maximum decline rates have a faster rise than would be expected from a single-parameter family of light curves, indicating that SN~Ia light curves are not a single-parameter family of varying widths. Comparison of the $B$-band rise-time distribution for spectroscopically normal SNe~Ia to those exhibiting high-velocity spectral features indicates that high-velocity (HV) SNe~Ia have shorter $B$-band rise times compared to their spectroscopically normal counterparts.  After normalising the $B$-band  light curves to \dmb\ = 1.1~mag (i.e., correcting the post-maximum decline to have the same shape as our template), we find that spectroscopically normal SNe~Ia have a rise time of $18.03 \pm 0.24~{\rm d}$, while HV SNe have a faster $B$-band rise time of $16.63 \pm 0.29~{\rm d}$. Despite differences in the $B$ band, we find that HV and normal SNe~Ia have similar rise times in the $V$ band. Furthermore, the initial rise of a SN~Ia $B$-band light curve follows a power law with index  $2.20^{+0.27}_{-0.19}$,  consistent with a parabolic rise in flux predicted by an expanding fireball toy model.  We compare our early-time $B$-band data to models for the predicted signature of companion interaction arising from the single-degenerate progenitor scenario. There is a substantial degree of degeneracy between the adopted power-law index of the SN light-curve template, the rise time, and the amount of shock emission required to match the data.
\end{abstract}

\begin{keywords}
supernovae: general
\end{keywords}

\section{Introduction}
Type Ia supernovae (SNe~Ia) are believed to be the result of a thermonuclear runaway explosion of a C/O white dwarf (WD) approaching the Chandrasekhar limit  (see \citealt{hillebrandt00} for a review). Explosive nucleosynthesis up to \nic releases $\sim 10^{51}{\rm ~erg}$, unbinding the progenitor. The subsequent light curve is powered by injection of energy from the radioactive decay of \nico; the $\gamma$ rays degrade to longer wavelengths as they diffuse out through the expanding ejecta. 

The well-established relationship between the light-curve width and the luminosity at peak brightness allows SNe~Ia  to be ``standardizable candles" at optical wavelengths \citep{phillips93}, and possibly almost standard candles in the infrared \citep{krisciunas04, wood-vasey08, folatelli10}. Application of the width-luminosity relation along with colour information to SNe~Ia over cosmological scales has led to the discovery of the accelerating expansion of the Universe (\citealt{riess98, perlmutter99}; see also \citealt{astier06, riess07, wood-vasey07, kowalski08, hicken09b,amunullah10}), indicating either the presence of ``dark energy'' having a negative pressure or a failure of general relativity on the largest scales. Recent work has also shown that including spectral flux ratios may also aid in reducing Hubble-diagram residuals \citep{bailey09, blondin11}.

Despite the successful cosmological application of SNe~Ia, the SN community lacks a clear understanding of the nature of their progenitor systems. Possible scenarios include a single-degenerate WD paired with a red-giant post-main-sequence star undergoing stable mass transfer until $M_{\rm WD}$ approaches $M_{\rm Ch} \approx 1.4$~\Msun \citep{whelan73,livio00}, a double-degenerate WD merger that reaches or surpasses $M_{\rm Ch}$ \citep{webbink84, iben94}, or the result of a sub-Chandrasekhar explosion of a WD steadily burning helium accreted from a companion \citep{shen09}. Each scenario carries with it tentative observational evidence and hardships imposed by unrealised theoretical predictions.

While considerable attention has been focused on the post-maximum decline of the SN~Ia light curve, less has been paid to the rise of the SN~Ia from explosion to maximum brightness. This is due to the dearth of data in the days following the SN explosion and the intrinsic difficulty of finding SNe~Ia shortly after explosion. The Lick Observatory Supernova Search (LOSS) has been successful at finding SNe in the nearby Universe (redshift $z < 0.05$) and the Sloan Digital Sky Survey (SDSS) Supernova Survey \citep{frieman08} has found and monitored 391 SNe at moderate redshift ($0.03 \le z \le 0.35$). Higher redshift searches such as the SuperNova Legacy Survey \citep{astier06} and ESSENCE \citep{wood-vasey07} benefit from cosmological time dilation, allowing the SN rise to be spread out over more days in the observer's frame.

A first attempt at quantifying SN~Ia rise times by \cite{pskovskii84} analysed 54 historical light curves gathered from photographic plates and visual $B$ magnitude estimates, finding a typical rise of 19--20~d. Recent attempts have made use of more reliable data taken with charge-coupled devices (CCDs) which are linear in translating photons to counts over a large dynamic range. A seminal paper by \citet[][hereafter R99]{riess99b} used observations of nearby SNe~Ia to find a $B$-band rise time of $19.5 \pm 0.2~{\rm d}$ for a normal SN~Ia having a decline of 1.1 mag between maximum light and 15~d after maximum in the $B$ band (i.e., $\dmb =1.1$ mag; Phillips 1993). Studies using data on higher redshift SNe~Ia have found a rise time in concordance with that of R99 \citep{aldering00,conley06}, advancing the notion that there is limited, if any, evolution in SN~Ia properties from high to low redshifts. Most recently, \cite[][hereafter C06]{conley06} found a rise of $19.10^{+0.18}_{-0.17} {\rm~(stat)} \pm 0.2 \rm{(sys)}$~d using SNLS data.

In an analysis of $B$- and $V$-band photometry of eight nearby SNe~Ia with excellent early-time data, \citet[][hereafter S07]{strovink07} found tentative evidence for two populations of $B$-band rise times. After correction for light-curve decline rate, S07 found that three SNe rise in $18.81 \pm 0.36$ d and five SNe rise in $16.64 \pm 0.21$ d. More recently, \citet[][hereafter H10]{hayden10a} analysed $B$- and $V$-band photometry 105 SNe~Ia from the SDSS sample and found that SNe~Ia come from a rather broad distribution of $B$-band rise times, with evidence indicating that slower declining events (e.g., more luminous SNe) have some of the fastest rise times. This result has significant implications for light-curve fitting techniques that rely on a single-parameter family of light curves and theoretical modeling of SNe~Ia. {H10 find an average $B$-band rise time of $17.38 \pm 0.17$~d, a departure from the results of R99 and C06 which H10 trace back to differences in fitting methods.

In this paper, we analyse available data on nearby SNe~Ia to measure the rise times, relying heavily on the recently released LOSS sample \citep{ganeshalingam10}. Previous analyses such as those of S07 and H10 use combined results from both $B$- and $V$-band photometry to measure the $B$-band rise time. In this paper, we compare the $B$-band rise time measured in the $B$ and $V$ bands and find that such a combination may not be appropriate. Instead, we present independent analysis of the $B$ and $V$ bands. Specifically, we define the rise time in a photometric band as the elapsed time between explosion and maximum light for that particular band. Nearby SNe offer the advantage of being able to be monitored by small-aperture telescopes and benefit from not requiring significant $K$-corrections, which at early times are ill defined because of a lack of available spectra to model the SN spectral energy distribution (SED).

\section{Data \label{s:data}}
\subsection{LOSS Light Curves}

LOSS is a transient survey utilizing the 0.76-m Katzman Automatic Imaging Telescope (KAIT) at Lick Observatory (\citealt{li00,filippenko01}; see also Filippenko, Li, \& Treffers 2011, in prep.). KAIT is a robotic telescope that monitors a sample of $\sim 15,000$ galaxies in the nearby Universe ($z < 0.05$) with the goal of finding transients within days of explosion. Fields are imaged every 3--10~d and compared automatically to archived template images, and potential new transients are flagged. These are subsequently examined by human image checkers, and the best candidates are reobserved the following night. Candidates that are present on two consecutive nights are reported to the community using the International Astronomical Union Circulars (IAUCs) and the Central Bureau of Electronic Telegrams (CBETs). Time allotted to our group on the Lick Observatory 3-m Shane telescope with the Kast double spectrograph \citep{miller93} is used to spectroscopically identify and study candidates. Between  first light in 1997 and 2010 September 30 UT, LOSS found over 865 SNe, 382 of which have been spectroscopically classified as SNe~Ia. The statistical power of the LOSS SNe is well demonstrated by the series of papers deriving the  nearby SN rates \citep{leaman11,li11a,li11b}.

In addition to the SN search, KAIT monitors active SNe of all types in broad-band \bvri filters. The first data release of \bvri light curves for 165 SNe~Ia along with details about the reduction procedure have been published by \cite{ganeshalingam10}. In summary, point-spread function (PSF) fitting photometry is performed on images from which the host galaxy has been subtracted using templates obtained $> 1$~yr after explosion. Photometry is transformed to the Landolt system \citep{landolt83,landolt92} using averaged colour terms determined over many photometric nights. Calibrations for each SN field are obtained on photometric nights with an average of 5 calibrations per field.

The LOSS light curves represent a homogeneous, well-sampled set of \bvri light curves. The average cadence is 3--4~d between observations, with a typical light curve having 22 epochs. Of the 165 \bvri\ light curves in the sample, 70 have data starting at least one week before maximum light.

\subsection{Light Curves from Other Nearby Samples}

In addition to the LOSS sample, here we include data from the following previously published SN~Ia samples: the Cal\'{a}n-Tololo sample \citep{hamuy96}, the Center for Astrophysics (CfA) Data Releases 1--3 \citep{riess99a,jha06,hicken09a}, and the Carnegie Supernova Project (CSP) dataset \citep{contreras09}. In cases where there are data from multiple samples, we chose the dataset with the best-sampled light curve to avoid introducing systematic calibration error. With the exception of the CSP sample, all light curves are in the Landolt photometric system. The CSP light curves are in the natural system of the Swope telescope at Las Campanas Observatory. Comparing LOSS $B$-band light curves (in the Landolt system) with CSP $B$-band light curves (in the Swope natural system), we find differences of $\sim 0.03$ mag which do not appear to be correlated with SN colour. We adopt 0.03 mag as the systematic uncertainty of the CSP Swope system light curves.

We also include individual light curves for SN 1990N \citep{lira98}, SN 1992A \citep{altavilla04}, SN 1994D \citep{patat96}, SN 1998aq \citep{riess05}, SN 1999ee \citep{stritzinger02}, SN 2003du \citep{stanishev07}, SN 2007gi \citep{zhang10}, and a preliminary reduction of SN 2009ig which was found by LOSS about 15~d before maximum light. In total, we have $BV$ light curves for 398 SNe.

\section{Methods\label{s:methods}}

In this section we detail the method used to measure the rise times of our sample. Most previous measurements of the SN~Ia rise time have used a single-stretch fit (see R99, C06). These studies ``stretched'' the template light curve along the time axis to determine the stretch value, $s$, that best fit the data. A light curve narrower than the template would have  $s < 1$, a wider light curve $s > 1$, and a light curve that matched the template perfectly $s = 1$. Explicit in this formalism is that a single stretch value applies to both the rising and falling portions of the light curve. Instead, we adopt a two-stretch fitting procedure first introduced by H10 to fit our $B$- and $V$-band data; we fit each band independently.}  Here we discuss our implementation of the two-stretch fitting routine with a template created mostly from LOSS data. 

\subsection{The Two-Stretch Fitting Method\label{s:two_stretch}}

In our two-stretch fitting routine, pre-maximum and post-maximum data are decoupled, allowing the two portions of the light curve to take on different stretch values to match a template. We define \sr\ to be the ``rise stretch'' and \sfl\ to be the ``fall stretch.'' The template has $s_{r} = s_{f} = 1.00$ by construction. The two stretched light-curve portions are joined at peak where the first derivative is 0, ensuring a continuous function with a continuous first derivative at maximum light.

Mathematically, the time axis is stretched such that 
\begin{equation}\label{e:tau}
\tau = \left\{
\begin{array}{lcr}
\frac{t - t_{0} }{s_{r} (1 + z)} & t \leq t_{0}  \\
\frac{t - t_{0} }{s_{f} (1 + z)} & t > t_{0},
\end{array}
\right.
\end{equation}
where $\tau$ represents the effective ``stretch-corrected" rest-frame epoch, $t_0$ is the time of maximum light, and $z$ is the SN redshift. 

Similar to other implementations of stretch \citep[e.g.,][]{goldhaber01}, we model each light curve as a function of time by
\begin{equation}
f(t) = f_{0} S(\tau),
\end{equation}
where $f_{0}$ is the peak flux and $S$ is our normalised template light curve. We perform a $\chi^2$ minimization to the quantity
\begin{equation}
\chi^{2} =\sum_{i}  \frac{ (F (t_{i}) - f(t_{i}))^{2}}{\sigma_{\rm {phot}}^2 + \sigma_{{\rm template}}^2},
\end{equation}
where $i$ is an index summed over all observations, $F(t_{i})$ are individual flux measurements, $\sigma_{{\rm phot}}$ is the photometric uncertainty including systematic error, and $\sigma_{{\rm template}}$ is the uncertainty in our template as described in \S \ref{s:template}. Our $\chi^2$ minimization fits for the values of $f_{0}$, $t_{0}$, $s_{r}$, and $s_{f}$. We restrict the fit to data within $\tau < +35 $ d relative to maximum light, after which SNe enter the nebular phase where the stretch parametrisation is no longer applicable \citep{goldhaber01}. 

The rest-frame rise time of a light curve (i.e., the elapsed time between explosion and maximum light in that band) is obtained by multiplying the measured $s_{r}$ by the fiducial rise time of the template. Following S07, we define the $B$-band fall time to be the amount of time required for the $B$-band light curve to decline by 1.1 mag starting at maximum light in $B$. We define the fall time for the $V$ band to be the required time for a $V$-band light curve to fall by 0.66 mag; the fall time of a light curve is $15\,s_{f}$~d. By construction, both templates with $s_{f} = 1$ have a fall time of $15~\rm{d}$. 

The $B$-band fall stretch \sfl\  is directly related to \dmb\ by the simple, empirical formula
\begin{equation}\label{e:dmb}
\Delta m_{15}(B) \approx 1.1 -1.70( s_{f}(B) -1.0) + 2.30 (s_{f}(B) - 1.0)^2,
\end{equation}
which was found by stretching our $B$-band template and reading off the resulting value
of \dmb\ mag. Similarly, for our $V$-band template, we find 
\begin{equation}\label{e:dmv}
\Delta m_{15}(V)  \approx 0.66 - 0.83 (s_{f}(V) - 1.0) + 0.94 (s_{f}(V) - 1.0)^{2}.
\end{equation}
Note that Equations \ref{e:dmb} and \ref{e:dmv} are only valid for the LOSS templates, and that $s_{f}(B)=1$ and $s_{f}(V) = 1$ respectively correspond to \dmb\ = 1.1 mag and \dmv\ = 0.66 mag.

 $K$-corrections are computed with the spectral series of \citet{hsiao07} which provides the spectral evolution of a SN~Ia with a one-day cadence. Using all available multi-colour photometry for a SN, the spectrum for the phase nearest the photometry epoch is warped to match the colours in the observer's frame using a third-order spline with knots placed at the effective wavelength of each available filter. $K$-corrections are computed from the warped spectrum and are typically $< 0.05$ mag.

For a single SN, our fitting procedure gives the date of maximum light, maximum flux, rise stretch, and fall stretch for both the $B$- and $V$-band light curves. For the purpose of comparing our results to those of H10, we compute the $B$-band rise time (i.e., the time between explosion and maximum light in $B$) found using both $B$- and $V$-band photometry. To determine the $B$-band rise time using $V$-band photometry, we measure the $V$-band rise time (the time between explosion and maximum light in $V$) and subtract off the time between maximum light in $V$ and maximum light in $B$. We employ this determination of the $B$-band rise time using $V$-band photometry to compare the rise-time behaviour of the two bandpasses over the same time period. Consequently, the determination of the $B$-band rise time using the $V$ band typically has larger uncertainties than using the $B$-band photometry, since we must also include the error from both the times of $B$ and $V$ maximum. We emphasise that the $V$-band rise time is defined to be the elapsed time between explosion and maximum light in the $V$ band.

\subsection{Template\label{s:template}}
The template plays an important role in measuring light-curve properties. A template which does not reflect the data will lead to fits with systematically incorrect measurements of light-curve properties \citep[H10;][]{aldering00}. Here we discuss the sample of objects and the method used to construct our light-curve templates for the $B$  and $V$ band. 

 We construct our templates from a sample of well-observed ``normal" SNe~Ia. Most of the objects come from the set of SNe~Ia observed by LOSS \citep{ganeshalingam10}, but we also include several objects published previously. SNe~Ia with excellent light curves but known peculiarities, such as SN 1999ac \citep{phillips06}, SN 2000cx \citep{li01:00cx, candia03}, SN 2002cx \citep{li03,jha06:02cx}, SN 2004dt \citep{leonard05:03du, wang06, altavilla07}, SN 2005hk \citep{phillips07}, and SN 2009dc \citep{yamanaka09, silverman10}, are avoided. The underluminous SN 1991bg-like objects show distinctly different photometric behaviour compared with the rest of the SN~Ia population, so they are excluded from the sample as well. In total, the set includes 60 objects, many of which are also in the  rise-time SN~Ia sample discussed in \S \ref{s:sample}.

For the zeroth-order template light curve, we adopt the fiducial $\Delta = 0$ (\dmb\ = 1.1 mag) template from the Multi-colour Light-Curve Shape (MLCS2k2) fitter \citep{jha07}.\footnote{Downloaded from www.physics.rutgers.edu/$\sim$saurabh/mlcs2k2/ on October 7, 2010.} Light-curve data are then fit using a two-stretch fitting parametrisation. The parameters being fit via a $\chi^2$ minimization technique are the time of maximum light $t_{0}$, the flux at maximum light $f_{0}$, the rise stretch $s_{r}$, and the fall stretch $s_{f}$ (see \S \ref{s:two_stretch} for details)  after correcting the light curves for time dilation using redshifts obtained from the NASA/IPAC Extragalactic  Database.\footnote{http://nedwww.ipac.caltech.edu/} The light-curve data are then normalised using the best-fit parameters to have a peak flux of 1.0 at $\tau = 0$ and de-stretched, using $s_{r}$ and $s_{f}$, along the time axis to match the template. After normalizing and de-stretching all of our light-curve data to match the shape of the template, we study the mean residuals between the data and the template.
 
While we find that even though the $\Delta = 0$ templates do a reasonable job in fitting the data, there are still small systematic trends in the fit residuals. For each band, we fit a smooth curve to the residuals and use it as a correction to the input template.  These ``refined" templates are then used to fit the light-curve data from our template sample again and the fit residuals are studied. This process is iterated until convergence is reached  --- that is, no systematic trend is observed in the fit residuals. Convergence is achieved within 5 iterations. We restrict this part of the template training procedure to data where the MLCS2k2 template is well defined, but before the SN enters the nebular phase. For the $B$ band this is within the range $-10 < \tau < +35$~d with respect to maximum light in $B$. For the $V$ band this is $-11< \tau < +35 $~d with respect to maximum light in $V$.

To estimate the uncertainties of our templates, we bin the fit residuals in 3~d intervals and calculate their root-mean square (RMS). Because a particular light-curve fit can have systematic residuals relative to the input template (i.e., several data points in one portion of the light curve all show negative residuals, while data in another portion all show positive residuals), the residuals at different epochs are correlated, and the RMS measurements are an overestimate of the true uncertainties. It is difficult to quantify how the residuals are correlated because different light-curve fits have different patterns of residuals. We assume that the residuals are  equally affected by the correlated errors in all portions of the template,  and apply a constant scaling factor to the RMS measurements so that the overall fit to all of the data has a reduced $\chi^2 \approx 1$.  These scaled RMS measurements are adopted as the uncertainties of our templates.

 Suffering from a dearth of early-time data,  we adopted an expanding fireball model to describe the SN light curve for $\tau < -10$~d relative to maximum light based on the arguments presented by R99 and employed in most rise-time studies thereafter. After explosion of the progenitor, the SN undergoes free, unimpeded  expansion such that its radius, $R$, expands proportionally with time, $\tau$.  Approximating the SN as a blackbody, the optical luminosity through a  broad-band filter on the Rayleigh-Jeans tail of the SED is given  by $L \propto R^{2}T \propto (v \tau)^2 T$. Despite recent observations indicating that $T$ may actually change substantially over this  period \citep{pastorello07,hayden10a}, if we assume that changes in $v$ and $T$  are modestly small, then $L \propto \tau^{2}$.  Wrapping our ignorance into  a ``nuisance parameter'' $\alpha$, we can write the flux in the rise-time region  as $f = \alpha(\tau + t_{r})^2$, where $t_{r}$ is the rise time of the template.
 
To determine the rise time for our template, we restrict our sample to light curves with data starting at $\tau \leq -10$~d relative to $B$-band maximum and at $\tau \leq -11$~d for the $V$ band. The following approach was adopted.

\begin{enumerate}
\item{Create a random realization of the light curves in our sample using reported photometry uncertainties for each data point and systematic calibration uncertainties for each light curve in our template section (see \S \ref{s:uncertainties} for details on how this is implemented).}
\item{Perform the two-stretch fitting routine, restricting the fits to data within the range $-10 < \tau < 35$~d for $B$ or $-11 < \tau < 35$~d for $V$. The fit is restricted to this range to avoid imposing a shape in the region where we will fit for the rise time.}
\item{Normalise and two-stretch correct the light curves in the sample.}
\item{Fit a parabola to the ensemble of light-curve points with $\tau \leq -10$~d for $B$ or  $\tau \leq -11$~d for $V$ with the constraint that the parabola is continuous with the template.}
\end{enumerate}

 The procedure outlined above was performed 1000 times for the $B$- and $V$-band data independently. We find a best-fit rise time of $17.92 \pm 0.19$~d for the $B$ band and $19.12 \pm 0.19$~d for the $V$ band.

We also attempted to find the best rise time for our template using the approach outlined by H10, except on a finer grid. We tested templates with different rise times in the range $15 \leq  t_{r} \leq 21$~d in 0.1~d increments (as opposed to the 0.5~d increments of H10). For each template, light curves with data starting at least ten days before maximum light  were fit using the two-stretch method (see \S \ref{s:two_stretch} for more details). Unlike the previous procedure to determine the rise time, all data at $\tau < +35$~d are used in the fit. We calculated the $\chi^2$ statistic for all stretch-corrected data in the region $ \tau\leq -10$~d for each template. We fit a fourth-order polynomial to the curve of reduced $\chi^2$ as a function of $t_{r}$. The minimum of the polynomial is taken to be the rise time that best fits our data.

Performing a Monte Carlo simulation of this procedure to draw 500 unique realizations of our dataset, we find a best-fit rise time of $17.11 \pm 0.09$~d for our $B$-band template and  $18.07 \pm 0.11$~d for our $V$-band template. These rise times disagree  at the 4--5$\sigma$ level with what we found above. It is unclear why these two procedures find such different results for the best-fit rise time for our template. However, when comparing a $B$-band template with a rise time of $17.11$~d to the data after they have been two-stretch corrected using stretches found with fits restricted to $-10 < \tau +35$~d, we see a significant systematic trend for data $\tau < -13$~d. Similar results are found in a $V$-band comparison with a template rise time of  $18.07 \pm 0.11$~d. While the nature of the discrepancy eludes us, we use a $B$-band rise time of $17.92 \pm 0.19$~d and a $V$-band rise time of $19.12 \pm 0.19$~d  to avoid introducing any systematic error. As a precaution, we have run our analysis using both sets of template rise times and find that while some of the final numbers change, they change systematically (by $\sim 0.50$--0.75~d) and do not affect any of our qualitative results or final conclusions. In \S \ref{s:companion}, we discuss using different rise times in the context of searching for signatures of interaction with the companion star in the single-degenerate scenario.

 \begin{figure}
\begin{center}
\includegraphics[scale=.4]{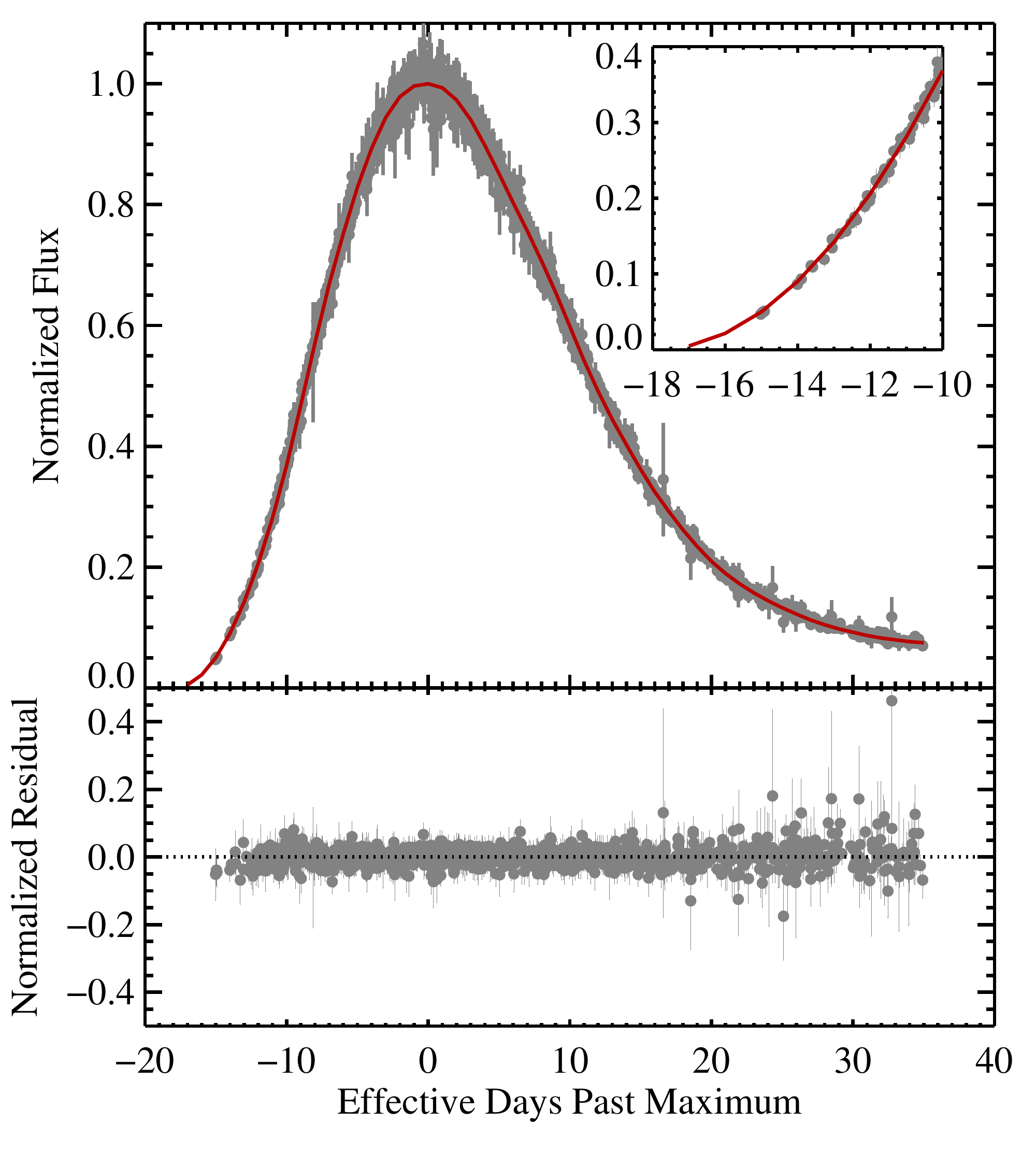}
\end{center}
{\caption{Stretch-corrected $B$-band light curves for \noo\ objects using the two-stretch method. All light curves have been shifted such that $\tau=0$~d is at maximum light and they have the same peak flux. Overplotted in red is the LOSS template. We restrict our fits to data within 35~d of maximum light. The inset panel is a detailed view of the early-time rise of the SN. The bottom panel shows the residuals between the stretch-corrected data and the template normalised by the template flux.
} \label{f:lc_two_stretch_corrected}}
\end{figure}

We caution the reader from interpreting the above template rise times as the ``typical'' rise time of a SN~Ia or even the rise time of a SN with $\Delta m_{15}(B) = 1.1$~mag. The rise time found above should be viewed as the rise time required to match the {\it shape} of our template light curve in the region $-10 < \tau < 0$~d. The rising portion ($\tau < 0$~d) of our template light curve based on the MLCS2k2 template does not {\it a priori} reflect a light curve with a decline $\Delta m_{15} (B) = 1.1~{\rm mag}$. Thus far, the goal of constructing our template was to find a light-curve shape that will fit our sample by applying independent stretches to the rising and falling portions of the light curve. To that extent, the actual number associated with the template rise time is meaningless; the significance is in the {\it shape} of the light curve. For example, we could construct an equivalent $B$-band template with a rise time of 35.84~d by stretching the rising portion ($\tau < 0$~d) of our 17.92~d template by a factor of 2. Using a template with a rise time of 35.84~d would decrease the measured rise stretches by a factor of 2. The final rise time for each SN found by multiplying the rise stretch by the fiducial rise time of the template will be the same for both templates. In \S \ref{s:rise_v_decline}, we will use the shapes of our template light curves to measure the rise and fall of our data sample and we will address the fall stretch-corrected rise time of SNe~Ia in our sample.

Figure \ref{f:lc_two_stretch_corrected} compares our template light curve to the stretch-corrected light-curve data. Residuals to the fit scaled by the template flux are plotted in the bottom panel of the figure. The template fits all portions of the light curve, without any systematic trends in the residuals.

\subsection{Estimating Uncertainties\label{s:uncertainties}}

To estimate the uncertainties in our fitting procedure, we use a Monte Carlo procedure including the effects of systematic calibration error from different photometric surveys, the uncertainty in rise time of our template, and the statistical photometric error. The prescription for one simulation for our Monte Carlo procedure is as follows.

\begin{enumerate}

\item For each survey, model calibration uncertainties by choosing a random photometric offset, and change all photometry from that survey by that random amount. The random offset is chosen for each survey assuming a Gaussian distribution with a mean offset of 0.0 mag and $\sigma = 0.03$ mag. SNe from the same survey will have the same photometric offset while SNe from a different survey may have a different offset.

\item Model the photometric error by perturbing every photometric point for each SN randomly based on its reported uncertainty, assuming the uncertainty is Gaussian.

\item Modify the LOSS template to have a rise time given by a random draw from a Gaussian distribution defined by the mean of 17.92~d and $\sigma = 0.19$~d for $B$ or 19.12~d and $\sigma = 0.19$~d for $V$ (see \S \ref{s:template} for how these values were derived). 

\item Fit simulated data with the modified template.

\end{enumerate}
We perform 1000 Monte Carlo simulations for our $B$- and $V$-band photometry independently. We take the parameter uncertainties to be the standard deviation in our 1000 trials and keep track of the covariance matrix between fit parameters.

\begin{figure}
\begin{center}
\includegraphics[scale=.45]{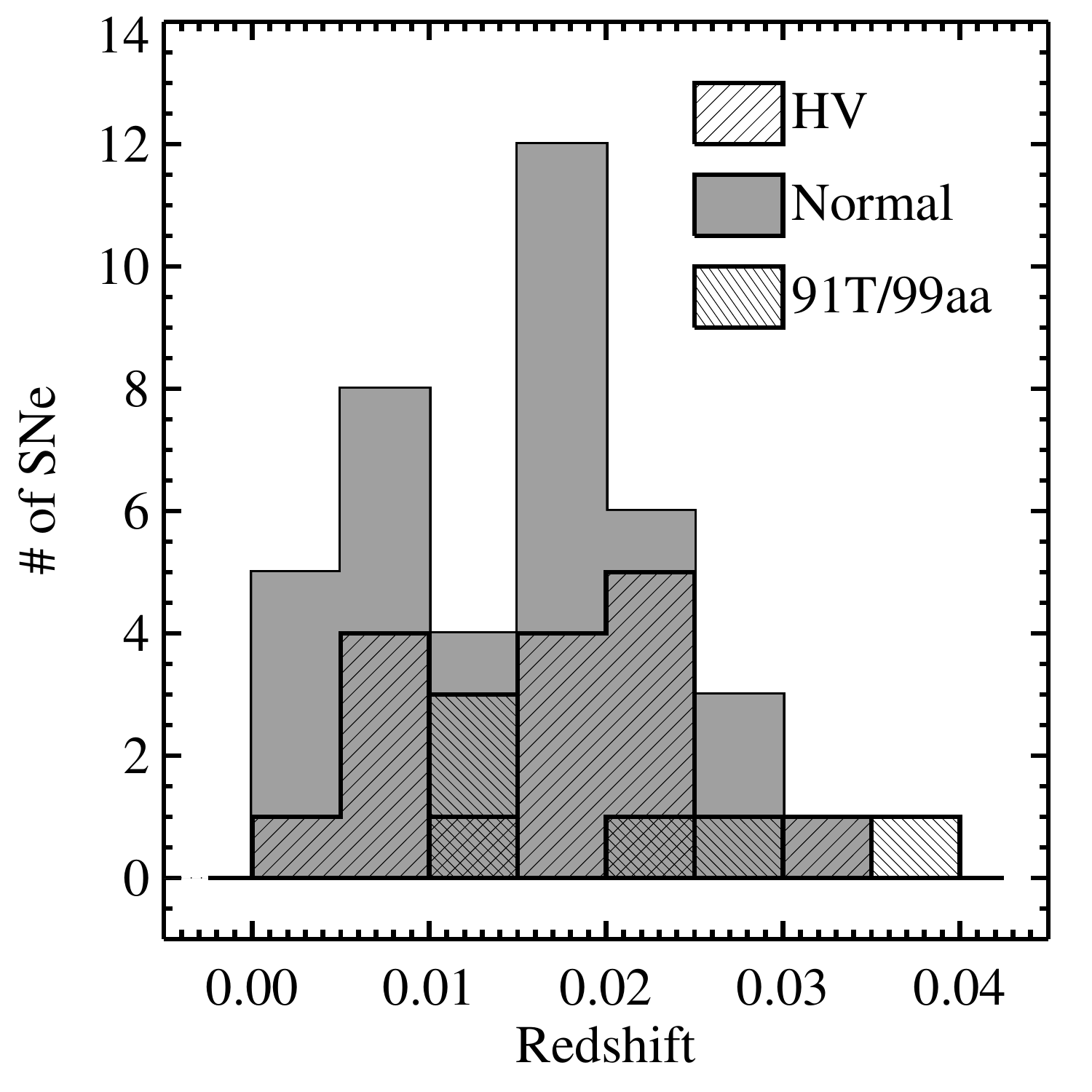}
\end{center}
{\caption{The redshift distribution of our sample broken into the three spectroscopic subclassifications: Normal, High Velocity, and SN 1991T/1999aa-like. The median redshift for the entire sample of \noo\ objects is 0.016.}\label{f:z_dis}}
\end{figure} 

\subsection{The Rise-Time Sample\label{s:sample}}
We restrict our final sample to objects that have $\sigma_{t_{r,f}} \leq 1.5$~d in both the $B$ and $V$ bands to ensure that we are using objects where the rise and fall of the light curve are being fit well. By the nature of our Monte Carlo procedure, each light curve does not have a single value of $\chi^{2}$ associated with it. We instead identify SN light curves which have a median reduced $\chi^{2} < 1.5$ over the 1000 trials. Fits are visually inspected to identify the poor ones, which are excluded from our sample. We also place the somewhat strict requirement that the SN have data starting 7 {\it effective} days before maximum (defined by Equation \ref{e:tau}) to anchor the measurement of \sr. We explore the effects of relaxing or tightening these requirements in \S \ref{s:cuts} and find that changing the requirements does not affect our final results.

 Objects similar to the subluminous SN 1991bg \citep{filippenko92:91bg, leibundgut93:91bg} were unable to be fit satisfactorily by our fitting routine. This is not surprising given the significant differences in light-curve shape and spectral evolution between the SN 1991bg-like SNe~Ia and ``Branch-normal SNe~Ia'' \citep{branch92}.  In \S \ref{s:91bg} we discuss our attempt to measure the rise times of SN 1991bg-like objects. We also exclude SN 2005hk, which is of the peculiar SN 2002cx subclass \citep{li03, jha06:02cx,phillips07}. 

Of the initial $BV$ light curves for 398 SNe, 95 have data starting at least 7 effective days before maximum light that could be fit with our two-stretch routine. Of the 95, 63 have good fits with rise and fall times that be can reasonably measured to within an uncertainty of 1.5~d. 

For the purposes of analysing the rise time of different spectroscopic subclasses, we further break our sample into three groups: Normal, High Velocity (HV), and SN 1991T/1999aa-like \citep{filippenko92:91t,li01b,garavini04}. For the classification of HV objects, we adopt the criterion of \cite{wang09} that the average $v_{\rm Si}$ for spectra taken within one week of maximum light is $\ga 1.2 \times 10^{4}$ km s$^{-1}$. Objects which have $v_{{\rm Si}} \le 10^{4}$ km s$^{-1}$ (more typical Si velocities for a normal SN~Ia) are classified as normal SNe~Ia.

\begin{figure}
\begin{center}
\includegraphics[scale=.4]{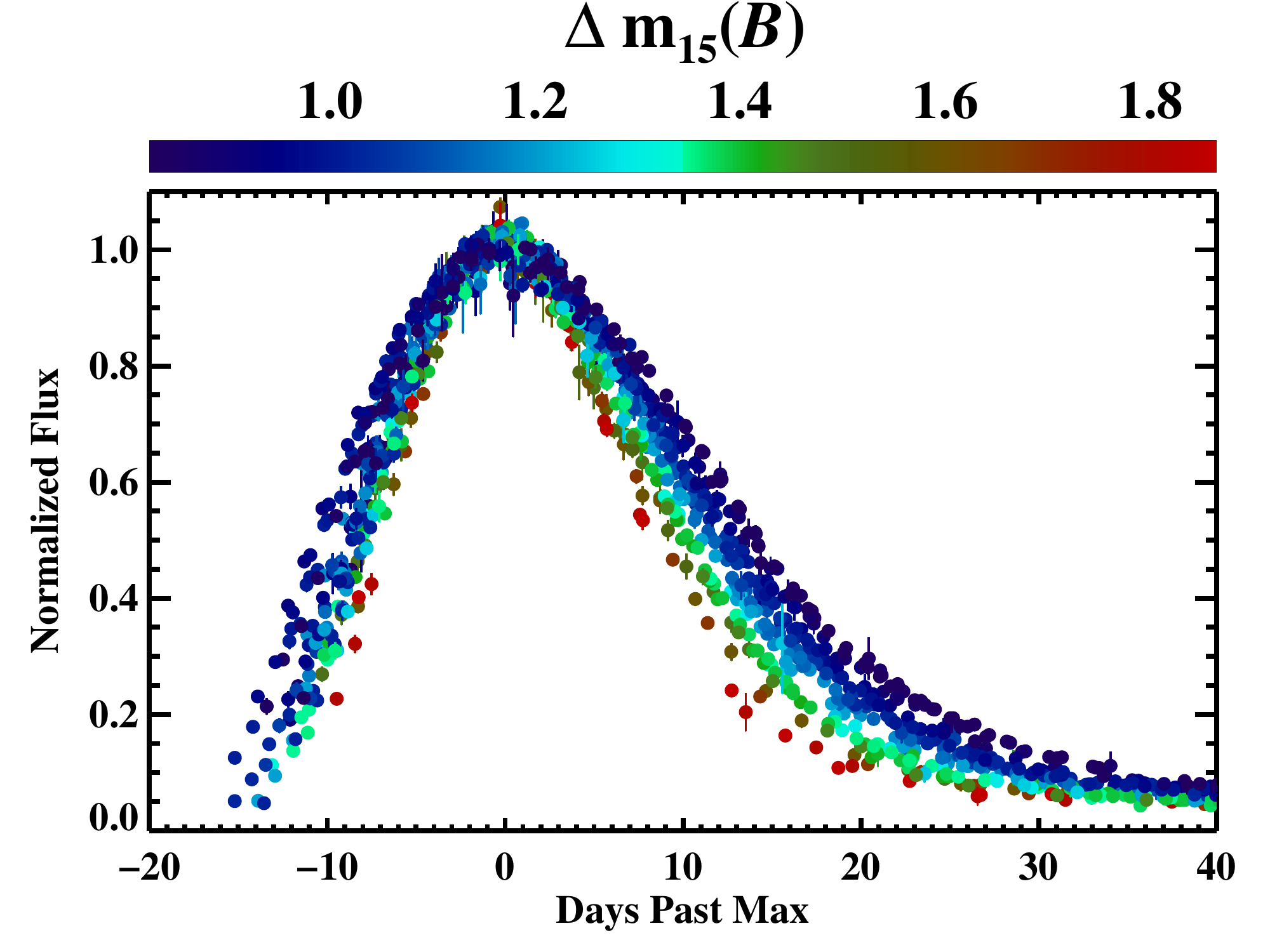}
\end{center}
{\caption{$B$-band light curves for the SNe used in this analysis. The light curves have been shifted such that they share the same peak flux and $t=0$ is the time of maximum light. The light curves are well sampled and start at least one week before maximum light. The SNe are coded by \dmb\ which measures the the post-maximum decline. The collection of light curves strongly suggests that SNe with slow rise also have a slow decline.
 }\label{f:all_rise_times}}
\end{figure}

We also spectroscopically identify SN 1991T/1999aa-like objects in our sample. The combination of SN 1991T-like and SN 1999aa-like objects is based off of previous studies which note the similar photometric and spectroscopic properties of the two subtypes \citep{ li01b, strolger02}. In particular, both exhibit broad light curves (\dmb $\approx 0.8$--0.9 mag) and spectra indicative of higher photospheric temperatures than normal SNe~Ia  (see \citealt{filippenko97} for a review of the spectroscopic diversity of SNe~Ia).

Our subclass information comes from spectra taken within one week of maximum light observed as part of the Berkeley SuperNova Ia Program (BSNIP; Silverman et al. 2011, in prep.). Using a modified version of the SuperNova IDentification (SNID) code \citep{blondin07}, we are able to classify SN spectra using a cross-correlation algorithm against a spectral database of know subtypes. Classification of SN 1991T/1999aa-like objects requires a spectrum within a week of maximum light to avoid confusion with normal SNe~Ia \citep{li01b}.

In instances where a SN in our sample has a classification in \cite{wang09}, we adopt their subclassifications. Otherwise, our subclassifications are based on BSNIP spectra. We are left without subclass identifications for only SNe 1992bo and 1992bc; hence, these two objects are excluded from our subsequent analysis. Our final sample consists of \noo\ SNe, including 39 normal, 16 HV, and 6 SN 1991T/1999aa-like SNe. Figure \ref{f:z_dis} shows the redshift distribution of our sample for each of the three spectroscopic subclassifications.

We plot the $B$-band light curves for the \noo\ SNe that passed our cuts in Figure \ref{f:all_rise_times}. The SNe are shifted relative to the time of maximum light and normalised to have the same peak flux. Qualitatively, objects with slower post-maximum declines (i.e., smaller \dmb) have slower rise times. In \S 4.1, we explore this relationship in more depth.

H10 analyse a sample of 41 nearby SNe~Ia drawn from S07, R99, and the CfA3 sample \citep{hicken09a}. Our analysis benefits from the addition of the LOSS SN~Ia sample which has $BV$ data starting at least a week before maximum light for 70 SNe. Data from the LOSS sample make up $\sim$70\% of the objects that pass our cuts.

 \begin{figure}
\begin{center}
\includegraphics[scale=.5]{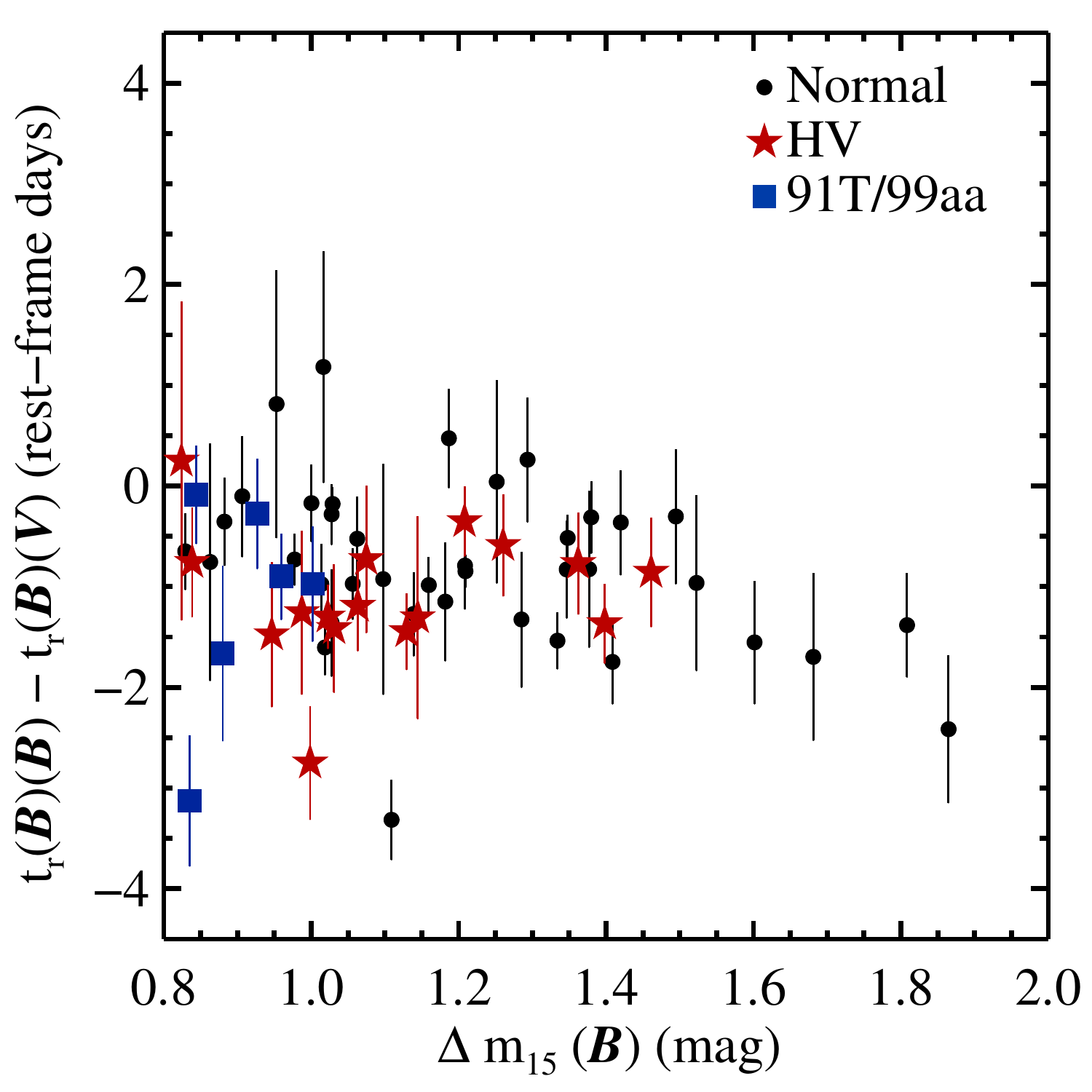}
\end{center}
{\caption{ Difference in the rest-frame $B$-band rise time derived from the $B$ band, $t_{r}(B)(B)$, and the $V$ band, $t_{r}(B)(V)$, as a function of \dmb. Differences are generally small, although there appears to be a small trend in the rise time as a function of \dmb. We find evidence for a systematic difference between the two measurements.} \label{f:b_v_rise}}
\end{figure}

\subsection{$B$- and $V$-Band Results: To Combine or Not to Combine?\label{s:b_v}}
In similar analyses of the $B$-band rise-time distribution, S07 and H10 combine stretches in the $B$ and $V$ bands using an error-weighted mean to produce a final single $B$-band rise time and fall time for each SN. H10 found evidence for a weak trend between the difference in stretch values for $B$ and $V$ as a function of \dmb\ using measurements for 105 SDSS-II SNe~Ia. 

In this paper, we take a different approach and independently measure the $B$-band and $V$-band rise times from the corresponding photometric data. However, we can use the results of our fitting routine to look for similar trends in the $B$-band rise time.  In \S \ref{ss:brise}, we compare the $B$-band rise time derived from the $B$ band to that derived from the $V$ band. The $B$-band rise time found using the $V$-band data is measured by taking the $V$-band rise time and subtracting the time between maximum light in the $V$ and $B$ bands.

Comparisons by H10 are done in stretch space, while ours are between the measured rise/fall time. H10 use a single template rise time of 16.5~d for both the $B$ and $V$ bands for fitting the rise stretch of their data light curves. Going from rise stretch to rise time only requires multiplying the measured rise stretch by the rise time of their fiducial template. Our fitting routine measures the rise and fall values of stretch using a Monte Carlo simulation which randomly chooses a template rise time based on the best-fit template rise time and its Gaussian uncertainty (see \S \ref{s:uncertainties} for details). The measured rise-stretch values are tied to the template rise time used for the fit, making the rise time a more appropriate quantity to compare.

\begin{figure}
\begin{center}
\includegraphics[scale=.5]{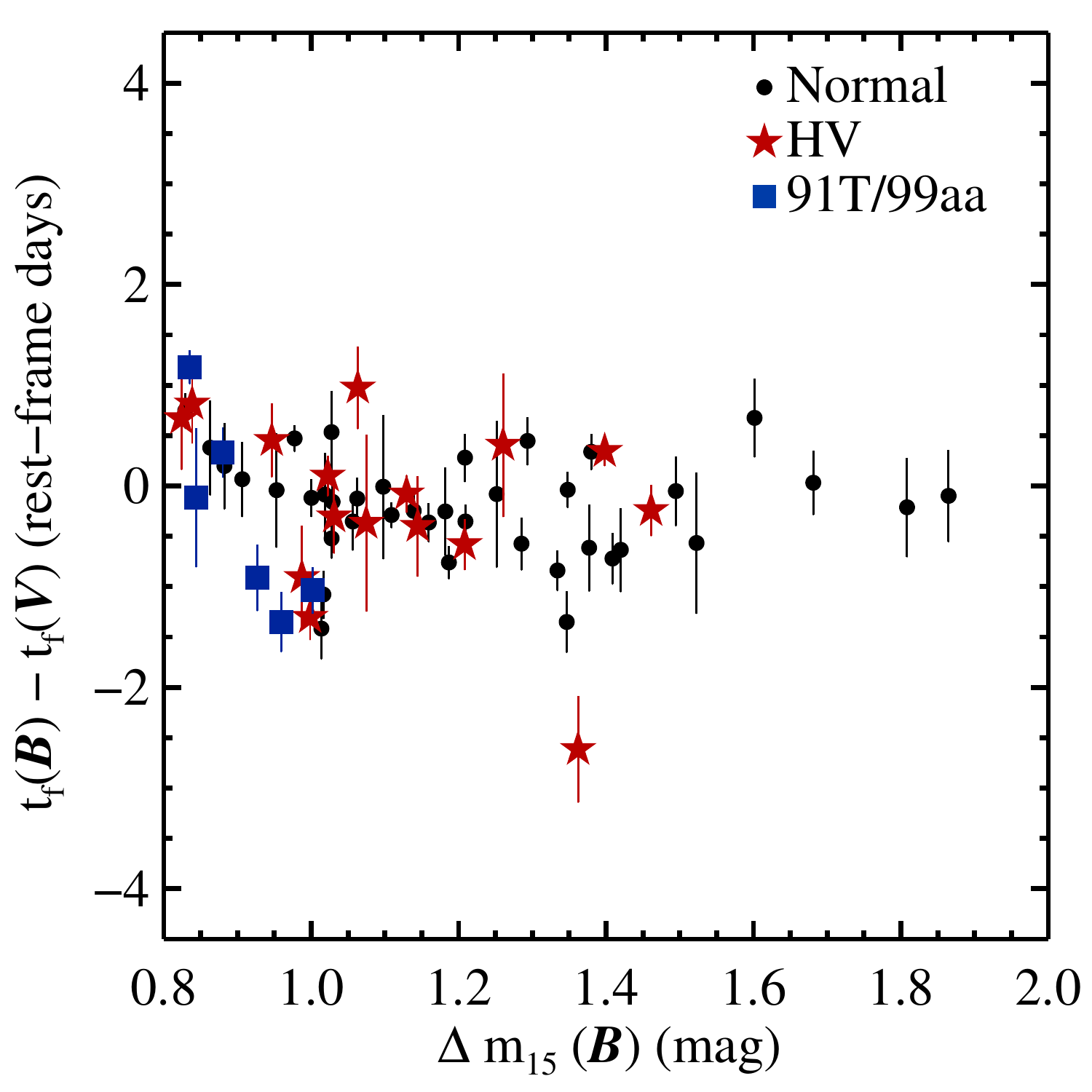}
\end{center}
{\caption{ Difference in the fall times in the $B$ band, $t_{f}(B)$, and the $V$ band, $t_{f}(V)$. The difference is consistent with 0~d. } \label{f:b_v_fall}}
\end{figure}

 \subsubsection{$B$-band Rise-Time Comparison\label{ss:brise}}
 
In Table \ref{t:diff_rise}, we show the differences in $B$-band rise time derived using the $B$ band, $t_{r}(B)(B)$, and the $V$ band, $t_{r}(B)(V)$, for our sample divided by spectroscopic subclassification.

 For our entire sample of \noo\ SNe, we find a mean difference of $-0.91\pm 0.10$~d (standard error of the mean), in the sense that the $B$-band rise time is shorter using $B$-band photometry compared to that derived using $V$-band photometry. Breaking our objects by spectroscopic subclassification, the difference in $B$-band rise time found using $B$- and $V$-band photometry is $-0.79 \pm 0.13$~d for normal SNe, $-1.08 \pm 0.16$~d for HV SNe, and $-1.17 \pm  0.45$~d for SN 1991T/1999a-like objects. We find evidence for a systematic difference of $\sim 1$~d between the 2 determinations of the $B$-band rise time.

In Figure \ref{f:b_v_rise} we plot difference in $B$-band rise time as a function of \dmb. There appears to be a small trend in the $B$-band rise time difference with increasing \dmb. Fitting a line results in a slope of $-0.54 \pm 0.44$~d mag$^{-1}$. The slope is computed by bootstrap resampling our sample to give 1000 realizations of our dataset. The mean and standard deviation of the distribution of the fit slopes are adopted as the most probable slope value and $1\sigma$ uncertainty. We do not find evidence of a significant trend.

 Our comparison shows evidence for a slight systematic trend that the rise time is $\sim 1$~d longer when measured with the $V$ band.  At least part of such a shift can be hidden by uncertainties in measuring the time of maximum flux of a light curve where the derivative slowly approaches 0 within a day of maximum light. Combining the two measurements will introduce systematic errors into an analysis.

\subsubsection {Fall-Time Comparison}

In a comparison of fall-time stretches derived from the $B$ and $V$ bands by H10, the authors found that objects with small \dmb\ had a larger $B$-band fall stretch than $V$-band fall stretch (i.e., $t_{f}(B) > t_{f}(V)$). We look for a similar trend in our data.

Recall that the $B$-band fall time is defined as the amount of time it takes for the $B$-band light curve to fall by 1.1 mag after maximum light in $B$, and the $V$-band fall time is defined as the amount of time it takes for the $V$-band light curve to fall by 0.66 mag after maximum light in $V$.  By construction, our templates both have a fall time of 15~d. 

 In Table \ref{t:diff_fall} we show the difference in fall times between the $B$ and $V$ bands (i.e., $t_{f}(B) - t_{f}(V)$). A comparison of the fall times for all \noo\ objects gives a mean difference of $-0.21 \pm   0.09$~d (standard error in the mean) between the $B$ and $V$ bands for all objects. There is no significant mean difference in the fall time measured between the $B$ and $V$ bands across all subclassifications.

  We plot the difference in fall times as a function of \dmb\ in Figure \ref{f:b_v_fall}. The mean fall-time difference is consistent with 0 d. Fitting a line to the data, we measure a slope of $-0.47 \pm 0.36$ d mag$^{-1}$, finding no significant evidence for a trend.

\begin{figure*}
\begin{center}
\includegraphics[scale=.7]{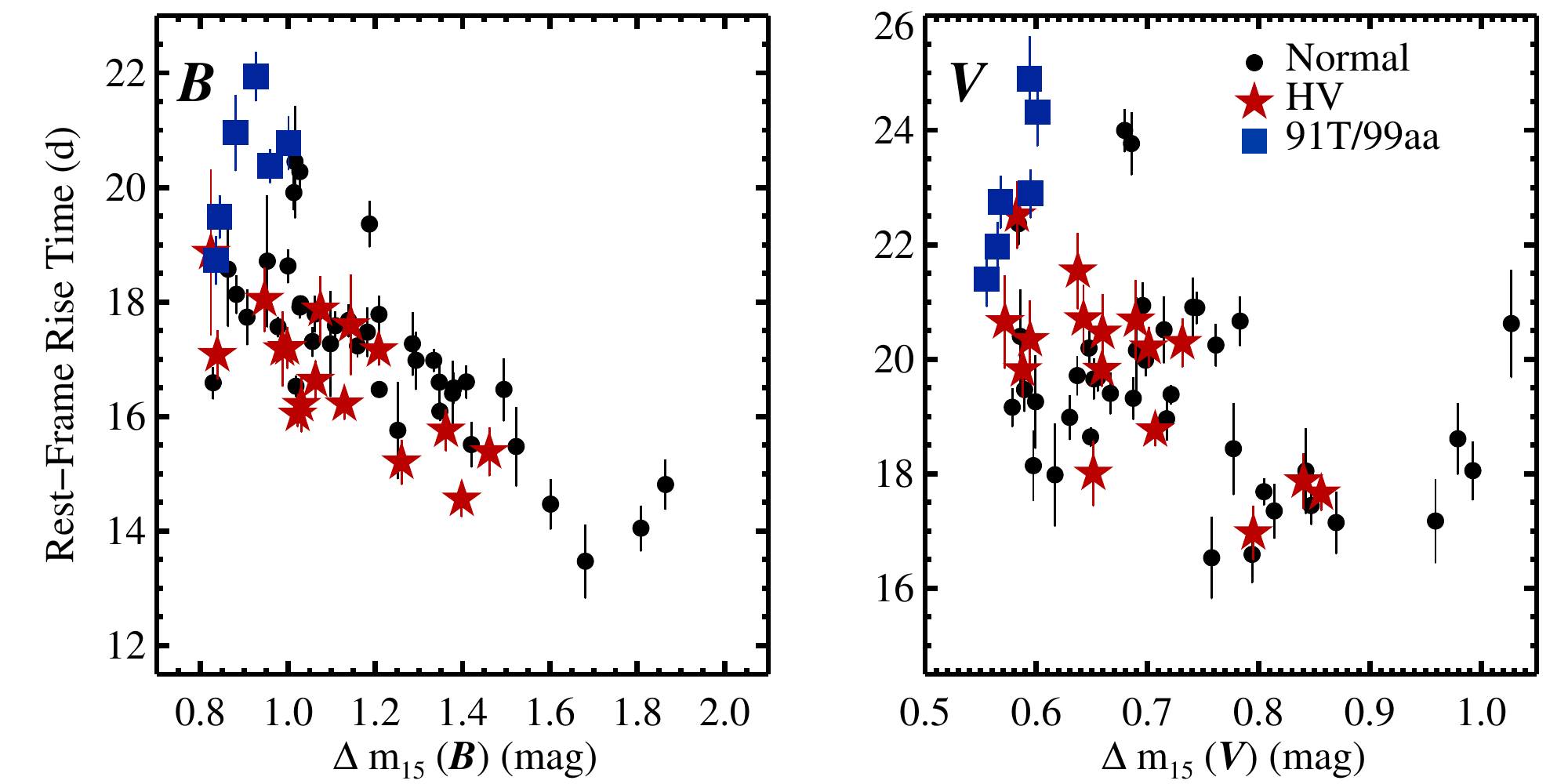}
\end{center}
{\caption{The rest-frame $B$-band rise time plotted as a function of \dmb\ (left), and the rest-frame $V$-band rise time as a function of \dmv\ (right). Note that the rise time is not stretch corrected. Using the two-stretch fitting method, we find a correlation between \dmb\ (calculated from $s_{f}(B)$ and Equation \ref{e:dmb}) and rise time. SNe with smaller \dmb\ (i.e., a slower post-maximum decline rate) have longer rise times. SNe that have been identified as HV are plotted as red stars. For fixed \dmb, HV SNe appear to have shorter rise times than their normal counterparts in the $B$ band. Overluminous SN 1991T/1999aa-like objects are plotted as blue squares. These objects have the smallest values of \dmb\ and the longest rise times. A less prominent, but similar correlation exists in \dmv\ (calculated from $s_{f}(V)$ and Equation \ref{e:dmv}) and the $V$-band rise time.}\label{f:m15b_v_tr}}
\end{figure*}

\begin{table}
\caption{Mean differences in $B$-band rise time derived using $B$- and $V$-band photometry by spectroscopic subclassification \label{t:diff_rise}}
\begin{center}
\begin{tabular}{lc}
\hline
Subclassification & $ t_{r}(B)(B) - t_{r}(B)(V)$  \\
\hline 
Normal                        &   $-0.79 \pm 0.13$ d \\
High Velocity             &   $-1.08 \pm 0.16$ d\\
SN 1991T/1999aa-like   & $-1.17 \pm 0.45$  d \\
All                                 &    $-0.91 \pm 0.10$ d\\
\hline
\end{tabular}
\end{center}
\end{table}

\begin{table}
\caption{Mean differences  in $B$- and $V$-band fall times by spectroscopic subclassification \label{t:diff_fall}}
\begin{center}
\begin{tabular}{lc}
\hline
Subclassification &  $ t_{f}(B) - t_{f}(V)$ \\
\hline 
Normal                        & $-0.20 \pm 0.08$  d\\
High Velocity             &    $-0.19 \pm 0.22$ d\\
SN 1991T/1999aa-like   & $-0.32 \pm 0.39$  d\\
All                                 & $-0.21 \pm 0.09$ d\\
\hline
\end{tabular}
\end{center}
\end{table}


\section{Analysis}
In this section we present our analysis of the $B$- and $V$-band rise-time distribution of our nearby sample. Given the results found in \S \ref{s:b_v}, we treat the $B$ and $V$ bands separately. In the following analysis, the rise time for a given band is defined as the elapsed time between explosion and maximum light in that band. Note that the results are not corrected for stretch unless indicated by a ${'}$ symbol, in which case the quantity is corrected for fall stretch.

\subsection{Rise Times Correlated with Decline \label{s:rise_v_decline}}
In Figure \ref{f:m15b_v_tr}, we plot the rest-frame $B$-band rise time as a function of \dmb\  and the rest-frame $V$-band rise time as a function of \dmv\ for our nearby sample of \noo\ objects. In both bands, there is a strong correlation between the rise time and the post-maximum decline, although the scatter indicates that the situation is more complicated than a simple one-to-one mapping between light-curve decline and the rise time. In general, SNe with slower post-maximum declines (e.g., small \dmb/ \dmv) have longer rise times. This correlation is evident in $B$, but shows more scatter in $V$. Using \noo\ objects, we calculate a Pearson correlation coefficient of $-0.69 \pm 0.03$ for $B$ and $-0.49 \pm 0.04$ for $V$. The probability of a Pearson coefficient $ < -0.49$ for two uncorrelated variables  and \noo\ measurements is $ < 0.01 \%$, indicating a highly significant correlation in both $B$ and $V$.   The correlation remains strong even when excluding SN 1991T/1999aa-like objects.

 Figure \ref{f:m15b_v_tr} indicates that SN~1991T/SN~1999aa-like SNe (i.e., overluminous SNe~Ia)  have the longest rise times. In contrast, H10 find that overluminous events have some of the shortest $B$-band rise times in their sample. However, H10 have few overluminous events and spectroscopic identifications were not included in their analysis. Instead, identifications of overluminous objects were based on \dmb.

\begin{table}
\caption{Rest-frame, fall-stretch corrected $B$- and $V$-band rise times\label{t:sc_rise}}
\begin{center}
\begin{tabular}{lcc}
\hline
Subclassification & $t_{r}{'} (B)$ &$ t_{r}{'} (V)$ \\
\hline 
Normal & $18.03 \pm 0.24$~d  & $20.23 \pm  0.44$~d \\
HV         & $16.63  \pm 0.29$~d & $19.43\pm  0.33$~d \\
SN 1991T/1999aa-like         & $18.05 \pm 0.69$~d  & $20.00 \pm 0.68$~d \\
 \hline
\end{tabular}
\end{center}
\end{table}

In Table \ref{t:sc_rise}, we give the fall-stretch corrected rise times, $t_{r}{'}$, for the various spectroscopic subclassifications in our sample for the $B$ and $V$ bands.  When $B$-band light curves are fall-stretch corrected to a light curve with $\Delta m_{15} (B) = 1.1$ mag (i.e., $s_{f} = 1$), the rise time of normal SNe~Ia and HV objects is (respectively) $18.03 \pm 0.24$~d (uncertainty in the mean)  and $16.63 \pm 0.29$~d, a $\sim 3 \sigma$ difference. H10 found an average $B$-band rise of $17.38 \pm 0.17$~d for the SDSS sample, consistent to $2.2 \sigma$ with our rise time for normal SNe~Ia.

 When we correct our $V$-band rise times to $\Delta m_{15}(V) = 0.66$~mag by dividing by the $V$-band fall stretch for normal objects, we find a fall-stretch corrected, rest-frame $V$-band rise time of $20.23 \pm 0.44$~d (uncertainty in the mean).  The fall-stretch corrected rise time of HV objects is $19.43 \pm 0.33$~d, consistent with the $V$-band rise time of normal objects. If we assume that a normal SN~Ia with $\Delta m_{15}(B) = 1.1$~mag corresponds to $\Delta m_{15}(V) = 0.66$~mag, then the fall-stretch corrected $V$-band rise time is $2.20 \pm 0.50$~d longer than the $B$-band rise time. This is within $1.5\sigma$ of the 1.5~d difference reported by R99 despite our disagreement on the actual measured $B$-band and $V$-band rise times for a $\Delta m_{15}(B) = 1.1$~mag SN (R99 measure  $19.5 \pm 0.2$~d and $21 \pm 0.2$~d for their $B$- and $V$-band rise times, respectively). For the HV objects, we measure a fall-stretch corrected rise-time difference of $2.80 \pm 0.44$ d. The larger difference (in absolute terms) in the $B$- and $V$-band rise times for HV objects compared to normal objects indicates that HV objects have a faster rise in the $B$ band than in the $V$ band (compared to normal objects).

\subsection{The Rise Time of High Velocity SNe}

Plotted as red stars in Figure \ref{f:m15b_v_tr} are objects spectroscopically subclassified as HV, while normal SNe~Ia are plotted as filled circles. In the $B$ band, HV objects appear to lie along a locus of points below that of normal SNe. For a fixed \dmb\ in the $B$ band, HV objects have a shorter rise than normal SNe. Analysing smaller samples,  \cite{pignata08} and \cite{zhang10} previously found that HV SNe~Ia appear to have a faster $B$-band rise for a given \dmb. This result is not evident in the $V$ band, where HV objects do not differ significantly from their normal counterparts.
 
The $\sim 3 \sigma$ difference in fall-stretch corrected rise time between HV and normal objects in $B$ and the lack of a significant difference in $V$ provides evidence for subtle differences in the photometric evolution of these two subclassifications. Recent work by \cite{wang09} and \cite{foley11} has shown that the two subclassifications have different $B_{\rm{max}} - V_{\rm{max}}$ pseudocolour\footnote{Note that $B_{\rm{max}} - V_{\rm{max}}$ is  the $B$-band magnitude at maximum light in the $B$ band minus the $V$-band magnitude at maximum light in the $V$ band. The dates of maximum light are typically offset by $\sim$2~d, making this quantity not an actual observed colour of the SN at a discrete time.} distributions, with HV objects typically having a redder pseudocolour. Furthermore, these studies have shown that separating HV and normal objects in cosmological analyses reduces the scatter in a Hubble diagram from 0.18 mag to 0.12 mag. 

\cite{foley11} provide a model which offers a possible explanation for why HV objects have a redder $B_{\rm{max}} - V_{\rm{max}}$ colour at maximum light. The two dominant sources of opacity in the atmosphere of a SN are electron scattering and line opacity from Fe-group elements. Electron scattering opacity is wavelength independent, while line opacity from Fe-group elements is most significant at wavelengths shorter than 4300 \AA~\citep[][hereafter KP07]{kasen07b}. The transition from electron scattering to Fe-group line opacity occurs near the peak of the $B$ band. SNe with high-velocity ejecta will have broader Fe-group absorption features, which will decrease flux in the $B$ band while having little effect on $V$-band flux.

 \cite{foley11} analyse models of an off-centre failed deflagration to detonation from KP07 to explore the expected differences in observables of HV and normal SNe.  The KP07 models studied a single SN with an off-centre ignition viewed from different angles. When viewed on the side nearest the ignition, the KP07 models produce a SN with HV features. When viewed from the side opposite the ignition, we will view a normal SN. Although the models presented in KP07 were intended to study the observational consequences of off-centre explosions, their model spectra and light curves conveniently produce a similar distribution of ejecta velocities to that observed between normal and HV SNe depending on viewing angle. The set of models predicts that HV objects should have redder colours at maximum light in comparison to normal objects (see Fig. 8 of \citealt{foley11}) on theoretical grounds.

Using the model light curves of KP07, we can explore how the rise time of HV objects will differ from that of normal objects.  If all other parameters are equal (i.e., Nickel mass, kinetic energy, etc.), the KP07 model predicts that HV SNe should have shorter rise times in $B$ compared to normal SNe (as indicated by the model light curves in their Fig. 11) due to the enhanced line opacity from Fe-group elements. Although not specifically stated by KP07, one would expect the $V$-band rise time of the two subclasses to be similar, since the opacity in this wavelength range is dominated by electron scattering.

Consequently, the KP07 models also predict that the enhanced opacity at blue wavelengths from HV components in the SN ejecta reduce the peak absolute $B$-band magnitude and hasten the evolution of the $B$-band light curve, producing a larger \dmb. This complicates a direct comparison between the rise time of HV and normal SNe and does not necessarily predict the observed result in the left panel of Figure \ref{f:m15b_v_tr}, that HV SNe lie on a locus of points below that of normal SNe in the $\Delta m_{15}(B)$--$t_{r}$ plane. One expectation of the KP07 models is a \dmb\ distribution for HV objects that is pushed to larger values in comparison to normal objects. We find a median \dmb = 1.09 mag for the HV objects and a median \dmb = 1.11 mag for the normal objects. However, our sample is most likely not representative of a complete sample of SNe~Ia and suffers from observational bias.

Without quantifying how HV ejecta change both \dmb\ and $t_{r}$, we cannot definitively claim that the models of KP07 explain our result. While it is beyond the scope of this paper to match theoretical models to our observations, the KP07 models offer a possible qualitative explanation for why there are significant differences between HV and normal SNe in the $B$ band, but not in the $V$ band. 

\begin{figure*}
\begin{center}
\includegraphics[scale=.7]{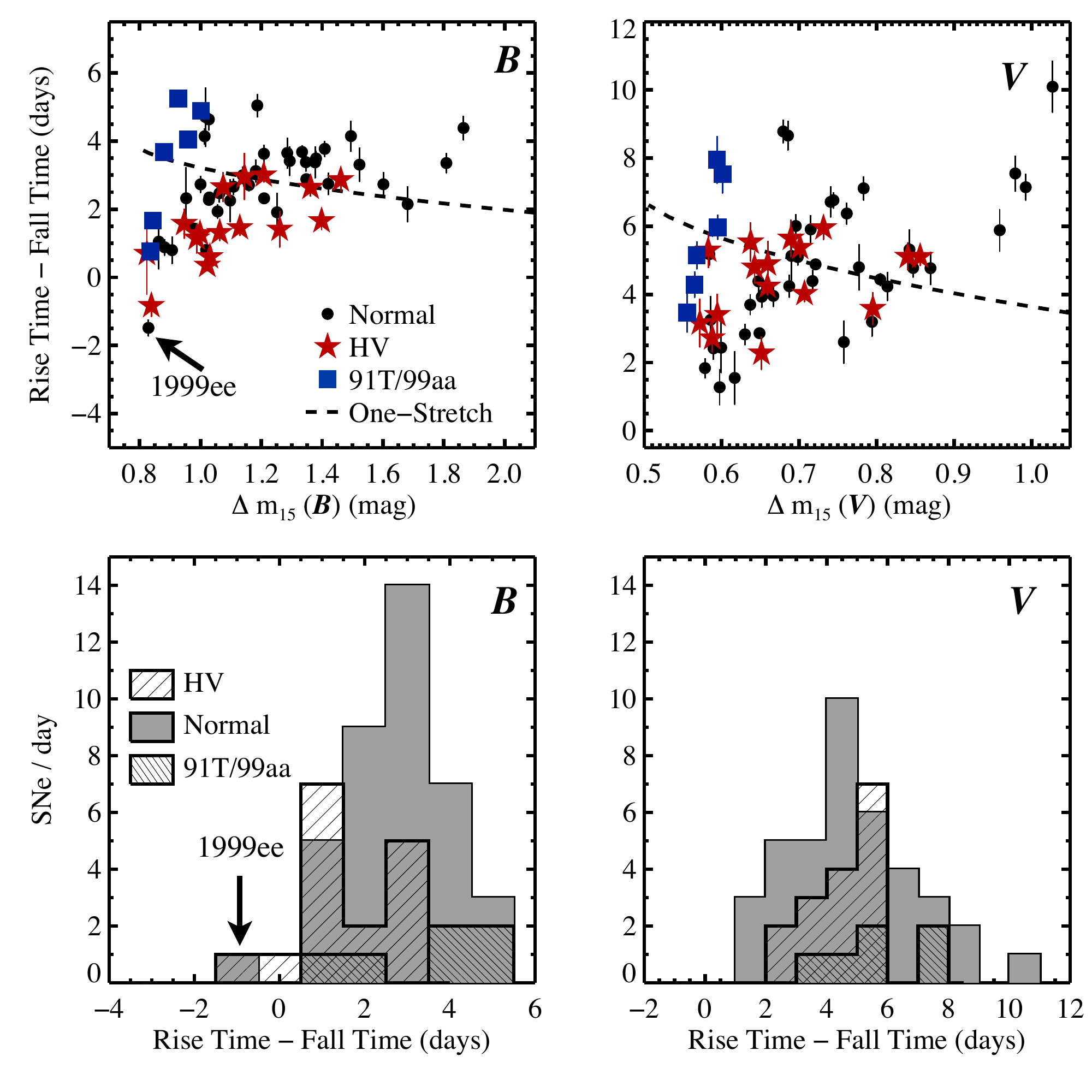}
\end{center}
{\caption{{\it Top panels:}  The rest-frame corrected rise time minus the fall time (RMF) as a function of \dmb\ for the $B$  band (left) and of \dmv\ for the  $V$ band (right). These quantities are not stretch corrected. Plotted as a dashed black line is the prediction using a single-stretch prescription where there is a one-to-one mapping of rise time to fall time. The nonlinearity in the prediction is due to the quadratic relationship between $s_{f}$ and \dmb\ in Equation \ref{e:dmb}. There is significant scatter about the line. We find that slowly declining SNe tend to have a faster rise than predicted by the single-stretch model. {\it Bottom panels:} The RMF distribution in the $B$ and $V$ bands. HV objects have a shorter $B$-band RMF in comparison to objects which are spectroscopically normal. A K-S test indicates a $< 0.01\%$ chance that the two distributions are drawn from the same parent distribution. No significant difference is seen in the $V$-band RMF distribution. SN 1999ee is the only normal SN with $B$-band RMF $< -1~\rm{d}$, although a previous study noted HV features in spectra of this object \protect \citep{mazzali05}. }\label{f:rmf_distribution}}
\end{figure*}

Recent work by \cite{maeda10} finds evidence that SNe with a high velocity gradient (HVG)
 in the \ion{Si}{II} line may be the natural result of viewing an asymmetric explosion. HVG SNe are defined by \cite{benetti05} as SNe that exhibit a time derivative in the velocity of the \ion{Si}{II} line $> 70$~km~s$^{-1}$~d$^{-1}$ around maximum light. \citeauthor{benetti05} provide evidence that the HV and HVG subclassifications are highly correlated. Looking at the \ion{Fe}{II} $\lambda$7155 and \ion{Ni}{II} $\lambda$7378 nebular emission in late-time spectra that trace the deflagration ashes, \citeauthor{maeda10} find that HVG SNe tend to exhibit redshifted lines while low-velocity gradient (LVG) SNe show a blueshift. The authors attribute this observational distinction to the difference between viewing an asymmetric explosion from the side nearest the site of initial deflagration (LVG) and the opposite side (HVG). The models of asymmetric explosions of SNe~Ia presented by \citet[][hereafter M11]{maeda11} indicate that viewing a SN from the far side, one expects HVG SNe to have longer bolometric rise times and smaller \dmb\ than comparable LVG SNe. This effect is more prominent in SNe with less $^{56}$Ni (for instance, compare models A0.3 and A0.6 in Fig. 11 of M11). The longer rise times in HVG objects is attributed to an increased optical depth for optical photons when viewing the explosion from the side opposite the explosion site (i.e., the site of \nic\ synthesis). 

This  is not necessarily inconsistent with our observational result that the HV SNe have a different rise-time distribution than normal SNe. If we view the SN from the side nearest the initial explosion site, we will see a LVG SN, and if we view the SN from the opposite side, we will see a HVG SN. Assuming HVG implies HV, we will measure a longer rise time and a smaller \dmb\  for a high-velocity object assuming the $B$ band roughly traces the bolometric behaviour.  This will move the HV object up and to the left of a normal SN in the \dmb--$t_{r}$ plane (Fig.~\ref{f:m15b_v_tr}). Depending on how the viewing angle affects both rise time and \dmb, the models of M11 could put the locus of HV points below that of normal SNe. However, the models presented in M11 do not outright predict a difference in rise times of HV objects in the $B$ and $V$ bands.

The models of KP07 and M11 offer opposite theoretical predictions for the rise time of HV SNe in comparison to normal SNe. KP07 predict that HV SNe should have larger \dmb\ and shorter rise times than normal SNe, while M11 predict smaller \dmb\ and longer rise times. The differences are rooted in the nature of the SN asymmetry (i.e., the distribution of intermediate-mass elements and $^{56}{\rm Ni}$) and the treatment of opacity. KP07 use an expansion opacity formalism which sums over individual lines, while M11 use a frequency averaged gray opacity. A better test of the HV models may possibly be found by looking at the \dmb\ distributions of a complete SNe~Ia sample, since both models predict that that asymmetries should influence the measured \dmb. \cite{wang09} found the \dmb\ distribution of HV and normal SNe to be strikingly similar despite different $B_{\rm max} - V_{\rm max}$ distributions, although it is not clear that their sample is complete. 

Despite evidence that HV objects have different rise-time properties than normal SNe, a clear physical picture remains elusive. Further efforts in modeling HV objects may shed light on the rise-time distribution of the different spectroscopic subclassifications.

\subsection{The Rise Minus Fall Distribution\label{s:rmf}}
In the top two panels of Figure \ref{f:rmf_distribution}, we compare the rise time minus the fall time (RMF) as a function of \dmb\ for the $B$ band (left panel) and \dmv\ for the $V$ band (right panel).   Note that the RMF has not been stretch corrected. As in Figure \ref{f:m15b_v_tr}, blue squares refer to SN~1991T/SN~1999aa-like objects, red stars refer to HV SNe~Ia, and black circles refer to spectroscopically normal SNe~Ia. Overplotted in a broken line is the expectation from a one-to-one mapping of rise time to fall time using a single-stretch parametrisation.  For the $B$ band we use a fiducial 18.03~d rise time and for the $V$ band we use a rise time of 20.23~d based on the fall-corrected rise times found in \S \ref{s:rise_v_decline}. Clearly, our sample does not strictly follow a single-stretch parametrisation. Similar to the results of H10, a number of slowly declining objects have a faster rise time (i.e., a smaller RMF) than expected from a one-stretch parametrisation in both $B$ and $V$.  We reiterate that based on the results of Figure \ref{f:m15b_v_tr}, more luminous SNe have longer rise times than less luminous SNe; however, Figure \ref{f:rmf_distribution} indicates that more luminous SNe have faster rise times than expected based on a single-stretch parametrisation.

 In the bottom two panels of Figure \ref{f:rmf_distribution}, we plot the rest-frame $B$- and $V$-band distributions of RMF for the various subclass identifications. In $B$, HV objects have a mean RMF of  $1.55 \pm 0.27$~d (uncertainty in the mean)  in comparison to normal SNe that have an RMF of $2.77 \pm 0.20$~d. The $4 \sigma$ difference in mean RMF offers evidence that the two distributions may be drawn from different populations. The mean RMF for our sample by subclassification can be found in Table \ref{t:rmf}. Despite significant differences in the $B$ band, the mean $V$-band RMFs for the spectroscopic subclassifications are consistent with one another.

A Kolmogorov-Smirnoff (K-S) test finds that the two groups of SNe have  a $\sim 0.01\%$ probability of being drawn from the same $B$-band RMF distribution and a $\sim 44\%$ probability of being draw from the same $V$-band RMF distribution.  The lone SN with RMF $< -1$~d in $B$ is SN 1999ee. \cite{mazzali05} find evidence of high-velocity features in spectra taken before maximum light. However, we retain the classification of \cite{wang09} and regard SN 1999ee as being normal. 

Our application of the K-S test indicates that HV and normal objects have a high probability of being drawn from different populations. However, the K-S test does not reveal whether this difference is physical in origin or a reflection of observational bias.  For instance, the difference in RMF between the two populations may also be a result of different stretch distributions. Such differences cannot be disentangled without knowledge of the observational bias in each of the photometric surveys used in our sample. Future surveys with proper spectroscopic follow-up observations and understanding of biases will aid in exploring the difference in RMF distribution for HV and normal SNe.

Our sample also includes 6 objects which show spectroscopic similarities to the overluminous SN 1991T or SN 1999aa. Based on the $B$- and $V$-band distribution of RMF for SN 1991T/1999a-like objects, these objects have slightly larger RMFs compared to normal and HV objects, however, the mean of the 1991T/1999aa-like distribution is consistent with the normal objects within $1\sigma$.

S07 find evidence for two distinct RMF distributions using eight SNe. While we have presented evidence for distinct HV and normal populations, S07 only have a single HV object in their sample (SN 2002bo). The other seven SNe are classified as normal by the criterion of \cite{wang09}. We do not see evidence for two populations of rise times within our spectroscopically normal SNe.

Similar to the results of H10, we find that our sample does not follow the expectation from a one-stretch parametrisation of light-curve shape. This is especially evident in $V$, where the trend appears to go in the opposite direction predicted by a single-stretch fit.

\begin{table}
\caption{Mean rise minus fall times for different spectroscopic subclassifications\label{t:rmf}}
\begin{center}
\begin{tabular}{lcc}
\hline
Subclassification & RMF ($B$) & RMF($V$) \\
\hline 
Normal                      & $ 2.77 \pm 0.20$~d & $ 4.83 \pm 0.33$~d \\
High Velocity            &$ 1.55 \pm 0.27$~d & $ 4.45 \pm 0.28$~d \\
SN 1991T/1999aa-like  & $ 3.38 \pm 0.73$~d & $ 5.72 \pm 0.73$~d \\
 \hline
\end{tabular}
\end{center}
 Note -- Quantities have not been corrected for stretch.
\end{table}


\subsection{Rise-Time Power Law\label{s:power}}
Previously, we assumed the rise in flux took the form of a parabola, $n = 2$, based on physical arguments presented in \S 3.1. In this section, we fit for the functional form of the rising portion of the light curve ($\tau \leq -10~\rm{d}$) as a power law of the form $f = A(\tau + t_{r})^n$, where $t_{r}$ is the rise time. Allowing $n$ to vary, we perform a $\chi^{2}$ minimization to find the best-fit power law to our $B$-band photometry. However, unlike the analysis in previous sections, we restrict our two-stretch fitting procedure to $-10 < \tau < +35~\rm{d }$ in order to avoid imposing a shape on the region we plan to fit.  We find a best fit of  $n = 2.20^{+0.27}_{-0.19}$, consistent ($1\sigma$) with the expanding fireball modeled by a parabola. The uncertainty in the power-law index is found using a Monte Carlo simulation similar to that outlined in \S \ref{s:uncertainties} except modified to only fit within the region $-10 < \tau < 35$~d.

Our result is in agreement with that of C06, who find $n = 1.8 \pm 0.2$, and H10, who find $n = 1.80^{+0.23}_{-0.18}$. However, H10 show evidence of significant colour evolution during this period of light-curve evolution, challenging our assumption of modest temperature change. They find a linear $B - V$ colour evolution of 0.5 mag between 15~d and 9~d before maximum light, leading to a temporal dependence closer to $n \approx 4$ rather than the $n = 2$ predicted by an expanding fireball.

We check for similar evidence of colour evolution in the expanding fireball phase in our sample.  Using the colour curves for SNe in the range of $-15 <\tau < -9$~d relative to maximum light, we create a median $B - V$ colour curve similar to the analysis presented by H10. We divide the data into 1~d interval bins and take the median $B-V$ colour for each bin. Corrections for Milky Way extinction were made using the dust maps provided by \cite{schlegel98}. No corrections were attempted for possible host-galaxy extinction. Our median colour curve shows a small change of 0.05 mag over the 6~d interval. However, inspecting the $B-V$ colour curve for SN 2009ig in this range, we find a drop of 0.5 mag over the same interval, similar to what is reported by H10. This leads us to suspect that our median colour curve does not necessarily reflect the full sample, and at least in individual SNe, there can be significant colour change at $\tau < -9$~d. The discrepancy between the median colour curve and that of SN 2009ig may be a result of using noisy data in the earliest time bins for the median colour curve.

\begin{figure*}
\begin{center}
\includegraphics[scale=.5]{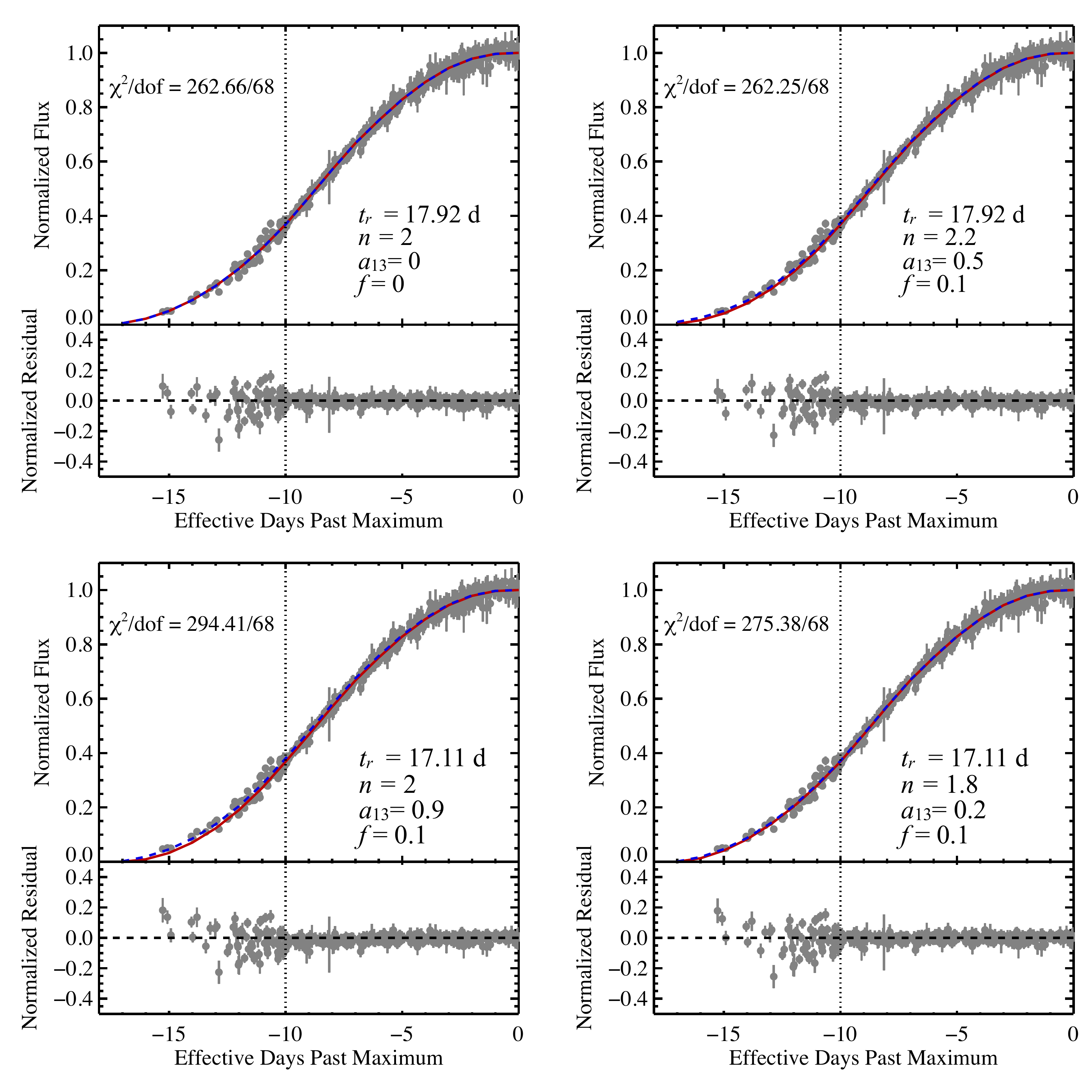}
\end{center}
{\caption{ Two-stretch corrected $B$-band light curves compared to the expected flux of a collision between the SN ejecta and a 1 \Msun companion assuming the companion is undergoing Roche-lobe overflow. Only data in the range $-10 < \tau < +35 $~d are used to stretch correct the light curves, to avoid imposing a shape onto data at $\tau \leq -10$~d. Plotted as a red line is the light-curve template with no shock interaction assuming various power laws for the initial rise, indicated by $n$, and rise times, $t_{r}$. Plotted as a blue dashed line is the expected ``shocked template," which is the expected shock emission added to the template assuming different separation distances $a_{13}$ and the fraction of SN explosions that show signs of interaction, $f$. The expected flux from the companion is estimated using the analytic models of \protect \cite{kasen10}. The bottom plot for each panel shows the resulting normalised residual curve between the data and the ``shocked template." We find a considerable degree of degeneracy between the adopted rise time, the power-law index for the initial rise, and the shock emission. As the four plots indicate, a similar minimum reduced $\chi^{2}$ can be achieved with $n=2$, $t_{r} = 17.92$~d and no shock emission (top left) or by varying the rise time, power-law index, and amount of shock emission (top right, bottom left, and bottom right, respectively). }\label{f:find_prog_emis}}
\end{figure*}

\subsection{Companion Interaction\label{s:companion}}
Recently, \cite{kasen10} proposed using emission produced from the collision of the SN ejecta with its companion star to probe the progenitor system.   Similar to the shock break-out in core-collapse SNe that is theoretically well understood \citep[e.g.,][]{klein78,matzner99} and observed \citep[e.g.,][]{soderberg08,modjaz09}, in the single-degenerate progenitor scenario the expanding SN ejecta are expected to collide with the extended envelope of the mass donating companion. In the case of a companion undergoing Roche-lobe overflow, the radius of the companion is on the same order of magnitude as the separation distance. The timing and luminosity emitted from the interaction will depend on the mass of the companion and the separation distance. The shock emission is expected to be brightest in the ultraviolet, but detectable in the $B$ band. \cite{hayden10b} looked for this signal in the rising portion of the $B$-band light curves of 108 SDSS SNe~Ia, finding no strong evidence of a shock signature in the data. Using simulated light curves produced with \verb+SNANA+ \citep{kessler09} and a Gaussian to model the shock interaction, the authors constrain the companion in the single-degenerate scenario to be less than a 6 \Msun\!\! main-sequence star, strongly disfavouring red giants (RGs) undergoing Roche-lobe overflow.

For this analysis, we focus on our $B$-band data where we have the best chance of detecting the signs of shock emission. \cite{kasen10} predicts that in the initial few days after explosion, the luminosity produced by the interaction with the companion will dominate the luminosity powered by $^{56}$Ni decay. Inspection of individual light-curve fits with the two-stretch fitting routine do not show the tell-tale signs of strong interaction. To take advantage of the power in numbers,  we analyse all of the light curves as an ensemble. We then compare data from the earliest light-curve epochs to the models of \cite{kasen10} to place constraints on the mass and distance to the companion.

We start by applying our two-stretch fitting routine again, but limiting the fit to data $ -10 < \tau < +35~\rm{d}$ in order to avoid forcing a shape on the earliest data and possibly suppressing the signature of interaction.  We construct models of ``shocked" template light curves including contributions from companion interaction using the analytic solutions for the properties of the emission found by \citet{kasen10}. Using equations for $L_{\rm{c,iso}}$ and $T_{\rm{eff}}$ from \cite{kasen10} (Eq. 22 and 25, respectively) and assuming the emission is that of a blackbody, we calculate the expected flux density at the peak of the $B$ band ($\lambda_{\rm{eff}} = 4450~\rm{\AA}$) at a distance of 10~pc. We normalise the flux density from the shock to a peak  SN~Ia magnitude of $M_{B} \ = -19.3~\rm{mag}$.

 Based on the opening angle of the shock interaction, \cite{kasen10} predicts that the interaction signature should be visible in $\sim 10$\% of SN~Ia explosions. Ideally, one should account for the viewing angle dependence of the detected shock emission. We make the simplification that we are either looking along the axis or we are not. Our light-curve models including the effects of interaction are the un-shocked template plus a fraction, $f$, of the collision flux calculated using the analytic model of \cite{kasen10}. A model with $f=0$ represents a situation in which shock emission is never detectable (and reduces to the unshocked template) and $f=1$ is a scenario in which shock emission is detectable in every SN explosion.

For most of the analysis in this paper, we adopted an expanding fireball model for the early rise of the light curve ($\tau \leq -10$~d) which assumed a power-law index of $n=2$. Under this assumption, we found that our early-time data were best fit by a rise time of 17.92 d. However, as discussed in \S \ref{s:power}, the assumption that the initial rise of the light curve has $n = 2$ may be somewhat questionable given rather significant changes in SN colour at early phases. As a test of the degeneracies between the added shock emission and the assumptions that go into constructing an unshocked template, we consider a number of different templates with different rise times and power-law indices. In addition to our nominal template with $t_{r} = 17.92$~d and $n=2$, we also try templates with $t_{r} = 17.11$~d and $n=2$, $t_{r} = 17.92$~d and $n=2.2$, and $t_{r} = 17.11$~d and $n=1.8$.

The parameters free to vary in the model for shock interaction are $a_{13}$, the distance to the companion normalised to $10^{13}~\rm{cm}$, $M_c$, the companion mass, and $f$, the fraction of SN explosions that produce detectable shock emission. For this simple analysis  the probed mass is fixed at $M_{c} = 1~{\rm M}_{\odot}$, to explore RGs as a possible companion, and we set $f$ equal to either 0 (i.e., no emission is detected) or 0.1, the expected fraction of SNe with detectable shock emission based on the opening angle of the shock. For each of our four different unshocked templates, we calculate the  minimum $\chi^{2}$ statistic for models by varying  $a_{13}$ to find the best-fit shocked template. The fit is restricted to data within the range of $-16 \leq \tau \leq -10 $~d.

We note that the minimum reduced $\chi^{2}$ in all of our fits for $a_{13}$ exceeds 1 and is usually closer to 3--4. This is not completely unexpected since data at $\tau \leq -10$~d were not included in the two-stretch fitting procedure used to normalise the light curves, and this may induce correlated errors into the data at $\tau \leq -10$~d. We also suspect that the reported errors for the earliest photometry epochs may be underestimated. Furthermore, the scatter in the data points may even be the physical result of a distribution of companion masses, separation distances, and viewing angles contributing to the emitted flux from shock interaction. Unfortunately, we are not in a position to disentangle what is contributing to the scatter.

Our final results for each template are shown in Figure \ref{f:find_prog_emis}. Plotted in red are the unshocked templates and in blue are the shocked templates including some contribution of companion interaction. In each case of an unshocked template, we can find an acceptable fit with similar $\chi^2$ by varying the separation distance. There is a significant degree of degeneracy between the parameters, making the task of disentangling the true sign of companion interaction at this level extremely difficult. For instance, for our template with $n=2$ and $t_{r} = 17.92$~d, we do not see signs of any interaction. However, had we used a template with $n=2$ and $t_{r} = 17.11$~d, we would find that if the fraction of SNe that showed signs of shock interaction is $f = 0.1$, then $\chi^{2}$ is minimised with $a_{13} = 0.9$. Similarly, when varying $n$ and $t_{r}$ in the unshocked template, suitable matches to the data can be found by adjusting the amount of shock interaction.

The degeneracy is partially broken by increasing the contribution of companion interaction to the ``shocked template" by either increasing $f$ or $a_{13}$.  Fixing $f = 0.1$ based on the expected opening angle of the shock, $a_{13} = 2$ increases the reduced $\chi^2$ by more than 1 for all of our unshocked templates.

In summary, we find that with our limited sample of early-time data, we cannot completely disentangle the behaviour  of the power-law index, the adopted rise time, and the amount of companion interaction.  This result makes use of all of the light-curve data at $\tau \leq -10$~d as an ensemble. Better constraints could be placed on individual, excellently sampled light curves in the rise region. In particular, efforts to obtain data at bluer wavelengths, where the shock interaction is stronger and more easily detectable, will greatly help to break degeneracies in the fitting process.

Recently, \cite{justham11} and \cite{stefano11} have argued that the transfer of angular momentum from the companion donor star to the C/O white dwarf acts to increase the critical mass required to explode and the time required for the white dwarf to explode. Consequently, the donor star may possibly evolve past the red giant phase, vastly reducing the cross section for interaction with SN ejecta. The subsequent shock emission for a RG would decrease beyond current detection limits, thus saving RGs as a possible donor in the single-degenerate scenario. The increased time to explosion allows the mass ejected from the envelope of the donor star time to diffuse to the density of the ambient interstellar medium, offering an explanation for the tight constraints  placed on the presence of $\rm{H}\alpha$ in nebular spectra of SNe~Ia \citep[][and references therein]{leonard07}.

\begin{figure}
\begin{center}
\includegraphics[scale=.42]{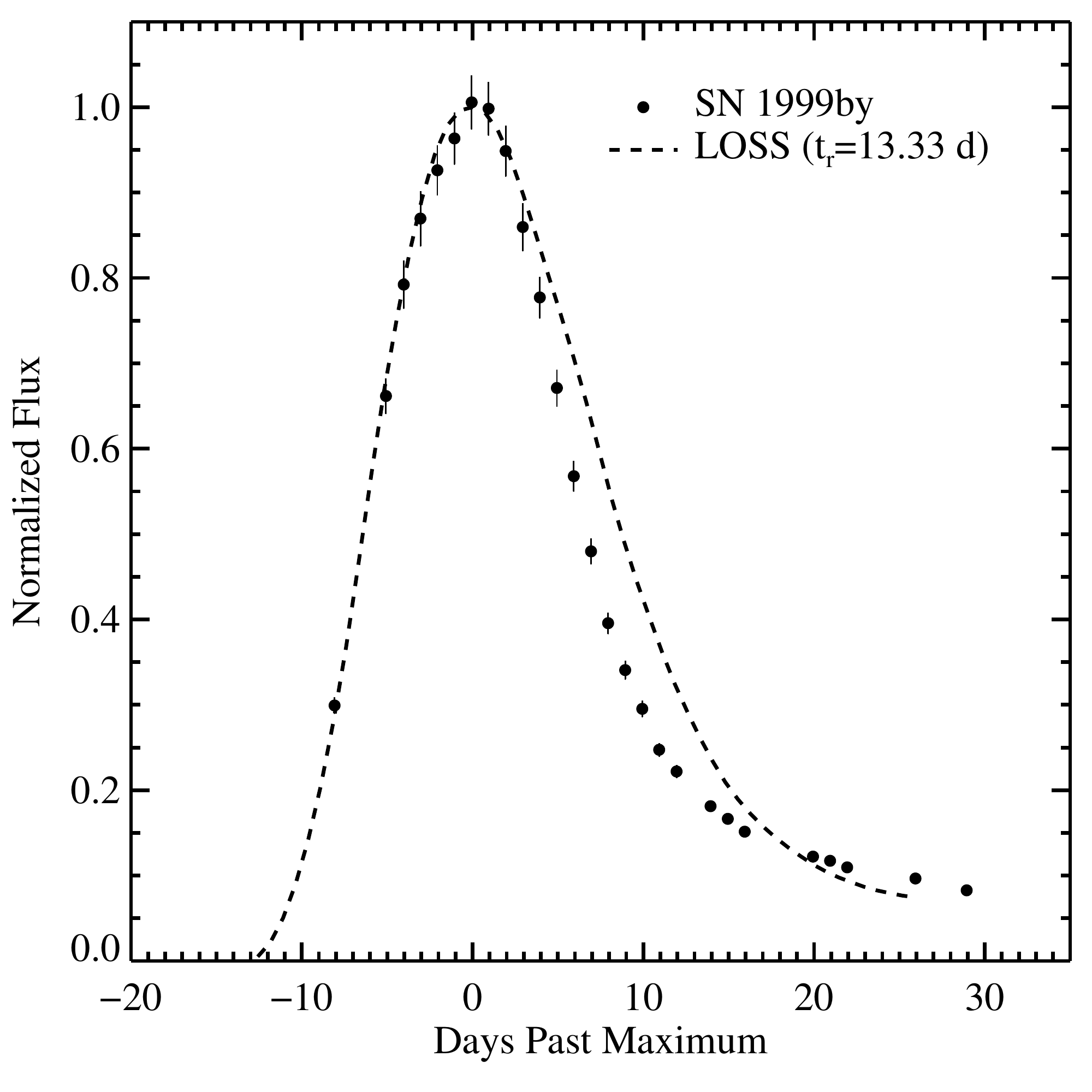}
\end{center}
{\caption{ Comparison of the $B$-band light curve for the SN 1991bg-like SN 1999by (solid circles) and the LOSS template (black dashed line) stretched along the time axis by 0.74 to match the rise portion of the light curve. The light-curve evolution of SN 1999by for $t > 10$~d past maximum makes it impossible to find a reasonable fit using our two-stretch fitting routine. If we restrict our fit to the pre-maximum portion of the light curve, we find a rise time of $13.33 \pm 0.40$~d.}\label{f:sn1991bg_fail}}
\end{figure}

\subsection{SN 1991bg-Like Objects\label{s:91bg}}

The analysis presented thus far has excluded SN 1991bg-like objects due to fits which had unacceptably large $\chi^{2}$ per degree of freedom. To investigate what was causing the poor fits, we focus on the $B$-band light curve of the SN 1991bg-like SN 1999by from the LOSS sample, which has data starting about $-10$~d before maximum light in the $B$ band. As seen in Figure \ref{f:sn1991bg_fail}, the largest difference in light-curve shape  between SN 1999by and our template occurs at $t > +10$~d, where the light curve of SN 1999by transitions to a slower linear decline not seen until $t >  +30$~d in normal SNe~Ia. Similar differences in light-curve shape for SN 1991bg are found by \cite{filippenko92:91bg} and \cite{leibundgut93:91bg}. The fit parameters of $s_{f}$  and $s_{r}$ are correlated; thus, a bad fit for $s_{f}$ will propagate into an incorrect determination of $s_{r}$. Given the different light-curve shapes for SN 1991bg-like objects and normal objects for $t > 10$ d, we are unable to acceptably fit $s_{f}$ and thus $s_{r}.$ For the purposes of exploring the rise time of SN 1991bg-like objects, we restrict our fit to the pre-maximum portion of the light curve using the date of maximum found with a low-order polynomial fit.   For SN 1999by, we find a best-fit rise time of $13.33 \pm 0.40$~d (overplotted as dashed lines in Fig. \ref{f:sn1991bg_fail}), indicating that they join other objects with large \dmb\ as the fastest risers in our sample. This matches the qualitative results found by \cite{modjaz01:98de} for the SN 1991bg-like SN 1998de.

\subsection{Impacts of Fitting Cuts\label{s:cuts}}

To ensure that the results described in the previous sections are not a result of the cuts made in  \S \ref{s:sample}, we reanalyse our data, both tightening and relaxing the constraints on the uncertainty in $t_{r}$ and $t_{f}$, $\sigma_{t_{r,f}}$. Most of the reported results are not highly sensitive to fitting cuts.  Restricting acceptable fits to reduced $\chi^{2} = 1.5$, a first epoch at $\tau = -10$~d, and $\sigma_{t_{r,f}} = 1$~d decreases the number of available objects to 26 normal, 9 HV, and 3 SN~1991T/SN~1999aa-like SNe~Ia. The probability that the $B$-band RMF populations are drawn from the same parent population increases to $\sim 4\%$.  The difference in the $B$-band fall-time corrected rise time, $t_{r}{'}$, between HV and normal SNe remains significant  at $-1.44 \pm 0.49$~d, indicating a faster rise for HV objects. Relaxing constraints to a reduced $\chi^{2} = 2$, first epoch at $\tau = -5$~d relative to maximum light, and  $\sigma_{t_{r,f}} = 1.5~\rm{d}$ increases the difference in fall-stretch corrected $B$-band rise time between HV and normal objects to $-1.86 \pm 0.49$~d. The $V$-band rise times remain consistent between HV and normal objects, as do the $V$-band RMF distributions. Overall, changing what we define as an ``acceptable" fit does not impact our results.


\section{Discussion}

We have presented an analysis of the rise-time distribution of nearby SNe~Ia in the $B$ and $V$ bands. Using a two-stretch fitting technique, we find that the SN rise time is correlated with the decline rate in the sense that SNe with broader light curves post-maximum (i.e., light curves with small \dmb /\dmv) have longer rise times.  While SN 1991bg-like objects could not be fit well by our two-stretch fitting procedure, we found that restricting our analysis to the pre-maximum data for the SN 1991bg-like SN 1999by gives a rise time of $13.33 \pm 0.40$~d. This follows the expected trend of a fast rise leading to a fast decline.

Using a sample of 105 SDSS SNe at intermediate redshifts ($0.037 \leq z \leq 0.230$), H10 find that there is a great diversity in $B$-band rise times for a fixed \dmb, and that the slowest declining SNe tend to have the fastest rise times. While we do see evidence for  scatter in rise times with \dmb, we find a strong correlation between \dmb\ and rise time in the sense that slowly declining SNe~Ia have longer rise times. This correlation is strong in the $B$ band, but weaker in the $V$ band.  The discrepancy with  H10 may be a result of combining $B$- and $V$-band stretches which was avoided in this analysis. 
 
We find that a single value of the stretch does not adequately describe the rising and falling portions of our light curves. Similar to the results in H10, more luminous SNe with longer fall times have shorter rise times than one would expect from a single-stretch prescription for light curves. This is especially evident in our analysis of $V$-band light curves. However, we reiterate that while luminous SNe have shorter rise times than expected from a single-stretch prescription, they still have longer rise times than less luminous SNe, contrary to the findings of H10.

 R99 and C06 found a fiducial $B$-band rise time of $\sim 19.5$~d for a ``typical" SN~Ia. In a departure from previous results, H10 report an average rise time of $17.38 \pm 0.18$~d. We find a fall-stretch corrected (i.e., corrected to have a post-maximum fall of \dmb\ = 1.1 mag) $B$-band rise time of $18.03 \pm 0.24$~d for spectroscopically normal SNe and $16.63 \pm 0.30$~d for HV SNe. Our $B$-band rise time for spectroscopically normal SNe with \dmb\ = 1.1 mag is in agreement with the average rise time found by H10 at the $2.2 \sigma$ level. When correcting our $V$-band light curves to a post-maximum fall of \dmv\ = 0.66 mag, we find a fall-stretch corrected $V$-band rise time of $20.23 \pm 0.44$~d.

After correcting for post-maximum decline rate, HV SNe~Ia have faster rise times than normal SNe~Ia in the $B$ band, but similar rise times in $V$. We find a $\sim 3\sigma$ difference in the fall-stretch corrected rise time between HV and normal SNe in the $B$ band.  The rise minus fall (RMF) distributions (not corrected for stretch) of the two subclassifications show significant differences in the two populations. The peak values of the distributions are offset by $1.22 \pm 0.34$~d in the $B$ band, with HV objects having a faster rise time. A K-S test indicates a $\sim 0.01$\% probability that HV SNe come from the same parent RMF population as normal SNe in the $B$ band. Despite differences in the $B$ band, we see no evidence of a difference in the RMF populations in the $V$ band.

Based on the model presented by \cite{foley11} and the models of KP07, we offer a possible qualitative explanation for why HV SNe should have a different $B$-band rise-time distribution than normal SNe, but similar $V$-band distributions. The physical origin of the difference is possibly rooted in the different opacity mechanisms at work in the $B$ and $V$ bands. Line blanketing from Fe-group elements is the dominant source of opacity for wavelengths shorter than $\sim 4300$ \AA, the peak of the $B$ band. At longer wavelengths, such as the $V$ band, electron scattering is the dominant source of opacity. Rapidly moving ejecta, as is the case with HV objects, will broaden absorption features, diminishing the $B$-band flux without affecting the $V$-band flux. Models from KP07 show that all other things being equal (e.g., Ni mass, kinetic energy), this leads to faster light curves with larger \dmb\ for HV objects. However, the enhanced opacity at short wavelengths also affects \dmb, complicating the application of the models to our result. Further modeling and observations of HV SNe~Ia are required to shed light on the photometric differences of this spectroscopic subclass. 

 We fit the earliest data in our sample ($\tau \leq -10~\rm{d}$) to find that the flux rises as a power law with index $2.20^{+0.27}_{-0.19}$. This is consistent with the unimpeded, free expansion of the expanding fireball toy model that predicts an index of 2. However, a preliminary analysis of SN~2009ig in the range $-15 < \tau < -9$~d shows evidence of significant colour evolution, contrary to the assumption of little to no colour evolution in the expanding fireball model. H10 find similar colour evolution in an analysis of the $B-V$ colour curve of SDSS SNe~Ia and derive an expected power-lax index of 4.

 We compare our early-time $B$-band data as an ensemble to models \citep{kasen10} of shock interaction produced from SN ejecta colliding with the mass-donating companion in the single-degenerate progenitor scenario. When relaxing our assumptions on the functional form of the early-time template light-curve behaviour (i.e., changing the rise time or power-law index of the rise), we find that our data require some amount of shock interaction to remove systematic trends. This indicates a level of degeneracy between the adopted template rise time, the power-law index, and the amount of shock interaction required to match the data. 

Future surveys with high-cadence search strategies will provide well-sampled SN light curves starting days after explosion, substantially adding to the sample of rise-time measurements. Complemented with spectroscopic follow-up observations, analysis of the RMF distribution can be further broken into different spectroscopic classes to provide insights into the underlying populations and the physics that differentiates them.

\section*{Acknowledgments}
We thank the Lick Observatory staff for their assistance with the operation of KAIT. We are grateful to the many students, postdocs, and other collaborators who have contributed to KAIT and LOSS over the past two decades, and to discussions concerning the results and SNe in general --- especially S. Bradley Cenko, Ryan Chornock, Ryan J. Foley, Saurabh W. Jha, Jesse Leaman, Maryam Modjaz, Dovi Poznanski, Frank J. D. Serduke, Jeffrey M. Silverman, Nathan Smith, Thea Steele, and Xiaofeng Wang. In addition, Silverman provided subclass identifications for SNe using unpublished data, while Daniel Kasen and Foley made insightful comments regarding high-velocity SNe.  We thank our referee, Alex Conley, for comments which significantly elevated the level of this manuscript. 

The research of A.V.F.'s supernova group at UC Berkeley has been generously supported by the US National Science Foundation (NSF; most recently through grants AST--0607485 and AST--0908886), the TABASGO Foundation, US Department of
Energy SciDAC grant DE-FC02-06ER41453, and US Department of Energy grant DE-FG02-08ER41563.  KAIT and its ongoing operation were made possible by donations from Sun Microsystems, Inc., the Hewlett-Packard Company, AutoScope Corporation, Lick Observatory, the NSF, the University of California, the Sylvia \& Jim Katzman Foundation, the Richard and Rhoda Goldman Fund, and the TABASGO Foundation. We give particular thanks to Russell M. Genet, who made KAIT possible with his initial special gift; Joseph S. Miller, who allowed KAIT to be placed at Lick Observatory and provided staff support; Jack Borde, who provided invaluable advice regarding the KAIT optics; Richard R. Treffers, KAIT's chief engineer; and the TABASGO Foundation, without which this work would not have been completed. We made use of the NASA/IPAC Extragalactic Database (NED), which is operated by the Jet Propulsion Laboratory, California Institute of Technology, under contract with NASA.

\label{lastpage}


\begin{thebibliography}{91}
\expandafter\ifx\csname natexlab\endcsname\relax\def\natexlab#1{#1}\fi

\bibitem[{{Aldering} {et~al.}(2000){Aldering}, {Knop}, \&
  {Nugent}}]{aldering00}
{Aldering} G., {Knop} R., {Nugent} P., 2000, \aj, 119, 2110

\bibitem[{{Altavilla} {et~al.}(2004)}]{altavilla04}
{Altavilla} G. {et~al.}, 2004, \mnras, 349, 1344

\bibitem[{{Altavilla} {et~al.}(2007)}]{altavilla07}
{Altavilla} G. {et~al.}, 2007, \aap, 475, 585

\bibitem[{{Amanullah} {et~al.}(2010)}]{amunullah10}
{Amanullah} R. {et~al.}, 2010, \apj, 716, 712

\bibitem[{{Astier} {et~al.}(2006)}]{astier06}
{Astier} P. {et~al.}, 2006, \aap, 447, 31

\bibitem[{{Bailey} {et~al.}(2009){Bailey}, {Aldering}, {Antilogus}, {Aragon},
  {Baltay}, {Bongard}, {Buton}, {Childress}, {Chotard}, {Copin}, {Gangler},
  {Loken}, {Nugent}, {Pain}, {Pecontal}, {Pereira}, {Perlmutter}, {Rabinowitz},
  {Rigaudier}, {Runge}, {Scalzo}, {Smadja}, {Swift}, {Tao}, {Thomas}, {Wu}, \&
  {The Nearby Supernova Factory}}]{bailey09}
{Bailey} S. {et al.}, 2009, \aap, 500, L17

\bibitem[{{Benetti} {et~al.}(2005){Benetti}, {Cappellaro}, {Mazzali},
  {Turatto}, {Altavilla}, {Bufano}, {Elias-Rosa}, {Kotak}, {Pignata}, {Salvo},
  \& {Stanishev}}]{benetti05}
{Benetti} S. {et al.}, 2005, \apj, 623, 1011

\bibitem[{{Blondin} {et al.}(2011){Blondin}, {Mandel}, \&
  {Kirshner}}]{blondin11}
{Blondin} S., {Mandel} K.~S., {Kirshner} R.~P., 2011, \aap, 526, paper ID A81

\bibitem[{{Blondin} \& {Tonry}(2007)}]{blondin07}
{Blondin} S., {Tonry} J.~L., 2007, \apj, 666, 1024

\bibitem[{{Branch} \& {Tammann}(1992)}]{branch92}
{Branch} D., {Tammann} G.~A., 1992, \araa, 30, 359

\bibitem[{{Candia} {et~al.}(2003)}]{candia03}
{Candia} P., {et~al.}, 2003, \pasp, 115, 277

\bibitem[{{Conley} {et~al.}(2006)}]{conley06}
{Conley} A. {et~al.}, 2006, \aj, 132, 1707 (C06)

\bibitem[{{Contreras} {et~al.}(2010)}]{contreras09}
{Contreras} C. {et~al.}, 2010, \aj, 139, 519

\bibitem[{{Di Stefano} {et~al.}(2011){Di Stefano}, {Voss}, \&
  {Claeys}}]{stefano11}
{Di Stefano} R., {Voss} R., {Claeys} J.~S.~W., 2011, ArXiv e-prints, 1102.4342

\bibitem[{{Filippenko}(1997)}]{filippenko97}
{Filippenko} A.~V., 1997, \araa, 35, 309

\bibitem[{{Filippenko} {et~al.}(2001){Filippenko}, {Li}, {Treffers}, \&
  {Modjaz}}]{filippenko01}
{Filippenko} A.~V., {Li} W.~D., {Treffers} R.~R., {Modjaz} M., 2001, in
   Small Telescope Astronomy on Global Scales, ed. {B.~Paczy\'{n}ski, W.-P.~Chen,
  \& C.~Lemme} (San Francisco: ASP, Vol. 246), 121

\bibitem[{{Filippenko} {et~al.}(1992{\natexlab{a}})}]{filippenko92:91t}
{Filippenko} A.~V. {et~al.}, 1992{\natexlab{a}}, \apjl, 384, L15

\bibitem[{{Filippenko} {et~al.}(1992{\natexlab{b}})}]{filippenko92:91bg}
{Filippenko} A.~V. {et~al.}, 1992{\natexlab{b}}, \aj, 104, 1543

\bibitem[{{Folatelli} {et~al.}(2010){Folatelli}, {Phillips}, {Burns},
  {Contreras}, {Hamuy}, {Freedman}, {Persson}, {Stritzinger}, {Suntzeff},
  {Krisciunas}, {Boldt}, {Gonz{\'a}lez}, {Krzeminski}, {Morrell}, {Roth},
  {Salgado}, {Madore}, {Murphy}, {Wyatt}, {Li}, {Filippenko}, \&
  {Miller}}]{folatelli10}
{Folatelli} G. {et al.}, 2010, \aj, 139, 120

\bibitem[{{Foley} \& {Kasen}(2011)}]{foley11}
{Foley} R.~J., {Kasen} D., 2011, \apj, 729, 55

\bibitem[{{Frieman} {et~al.}(2008)}]{frieman08}
{Frieman} J.~A. {et~al.}, 2008, \aj, 135, 338

\bibitem[{{Ganeshalingam} {et~al.}(2010)}]{ganeshalingam10}
{Ganeshalingam} M. {et~al.}, 2010, \apjs, 190, 418

\bibitem[{{Garavini} {et~al.}(2004)}]{garavini04}
{Garavini} G. {et~al.}, 2004, \aj, 128, 387

\bibitem[{{Goldhaber} {et~al.}(2001)}]{goldhaber01}
{Goldhaber} G. {et~al.}, 2001, \apj, 558, 359

\bibitem[{{Hamuy} {et~al.}(1996)}]{hamuy96}
{Hamuy} M. {et~al.}, 1996, \aj, 112, 2408

\bibitem[{{Hayden} {et~al.}(2010{\natexlab{a}})}]{hayden10a}
{Hayden} B.~T. {et~al.}, 2010{\natexlab{a}}, \apj, 712, 350 (H10)

\bibitem[{{Hayden} {et~al.}(2010{\natexlab{b}})}]{hayden10b}
{Hayden} B.~T. {et~al.}, 2010{\natexlab{b}}, \apj, 722, 1691

\bibitem[{{Hicken} {et~al.}(2009{\natexlab{b}}){Hicken}, {Wood-Vasey},
  {Blondin}, {Challis}, {Jha}, {Kelly}, {Rest}, \& {Kirshner}}]{hicken09b}
{Hicken} M., {Wood-Vasey} W.~M., {Blondin} S., {Challis} P., {Jha} S., {Kelly}
  P.~L., {Rest} A., {Kirshner} R.~P., 2009{\natexlab{b}}, \apj, 700, 1097

\bibitem[{{Hicken} {et~al.}(2009{\natexlab{a}})}]{hicken09a}
{Hicken} M., {et~al.}, 2009{\natexlab{a}}, \apj, 700, 331

\bibitem[{{Hillebrandt} \& {Niemeyer}(2000)}]{hillebrandt00}
{Hillebrandt} W., {Niemeyer} J.~C., 2000, \araa, 38, 191

\bibitem[{{Hsiao} {et~al.}(2007){Hsiao}, {Conley}, {Howell}, {Sullivan},
  {Pritchet}, {Carlberg}, {Nugent}, \& {Phillips}}]{hsiao07}
{Hsiao} E.~Y., {Conley} A., {Howell} D.~A., {Sullivan} M., {Pritchet} C.~J.,
  {Carlberg} R.~G., {Nugent} P.~E., {Phillips} M.~M., 2007, \apj, 663, 1187

\bibitem[{{Iben} \& {Tutukov}(1994)}]{iben94}
{Iben} I., Jr., {Tutukov} A.~V., 1994, \apj, 431, 264

\bibitem[{{Jha} {et~al.}(2007){Jha}, {Riess}, \& {Kirshner}}]{jha07}
{Jha} S., {Riess} A.~G., {Kirshner} R.~P., 2007, \apj, 659, 122

\bibitem[{{Jha} {et~al.}(2006{\natexlab{a}})}]{jha06}
{Jha} S., {et~al.}, 2006{\natexlab{a}}, \aj, 131, 527

\bibitem[{{Jha} {et~al.}(2006{\natexlab{b}}){Jha}, {Branch}, {Chornock},
  {Foley}, {Li}, {Swift}, {Casebeer}, \& {Filippenko}}]{jha06:02cx}
{Jha} S., {Branch} D., {Chornock} R., {Foley} R.~J., {Li} W., {Swift} B.~J.,
  {Casebeer} D., {Filippenko} A.~V., 2006{\natexlab{b}}, \aj, 132, 189

\bibitem[{{Justham}(2011)}]{justham11}
{Justham} S., 2011, \apjl, 730, L34

\bibitem[{{Kasen}(2010)}]{kasen10}
{Kasen} D., 2010, \apj, 708, 1025

\bibitem[{{Kasen} \& {Plewa}(2007)}]{kasen07b}
{Kasen} D., {Plewa} T., 2007, \apj, 662, 459 (KP07)

\bibitem[{{Kessler} {et~al.}(2009){Kessler}, {Bernstein}, {Cinabro}, {Dilday},
  {Frieman}, {Jha}, {Kuhlmann}, {Miknaitis}, {Sako}, {Taylor}, \&
  {Vanderplas}}]{kessler09}
{Kessler} R. {et al.}, 2009, \pasp, 121, 1028

\bibitem[{{Klein} \& {Chevalier}(1978)}]{klein78}
{Klein} R.~I., {Chevalier} R.~A., 1978, \apjl, 223, L109

\bibitem[{{Kowalski} {et~al.}(2008)}]{kowalski08}
{Kowalski} M. {et~al.}, 2008, \apj, 686, 749

\bibitem[{{Krisciunas} {et~al.}(2004)}]{krisciunas04}
{Krisciunas} K. {et~al.}, 2004, \aj, 128, 3034

\bibitem[{{Landolt}(1983)}]{landolt83}
{Landolt} A.~U., 1983, \aj, 88, 439

\bibitem[{{Landolt}(1992)}]{landolt92}
{Landolt} A.~U., 1992, \aj, 104, 340

\bibitem[{{Leaman} {et~al.}(2011)}]{leaman11}{Leaman} J.,{ Li} W., 
{ Chornock} R., \& {Filippenko} A.~V., 2011, \mnras, 412, 1419

\bibitem[{{Leibundgut} {et~al.}(1993)}]{leibundgut93:91bg}
{Leibundgut} B. {et~al.}, 1993, \aj, 105, 301

\bibitem[{{Leonard}(2007)}]{leonard07}
{Leonard} D.~C., 2007, \apj, 670, 1275

\bibitem[{{Leonard} {et~al.}(2005){Leonard}, {Li}, {Filippenko}, {Foley}, \&
  {Chornock}}]{leonard05:03du}
{Leonard} D.~C., {Li} W., {Filippenko} A.~V., {Foley} R.~J., {Chornock} R.,
  2005, \apj, 632, 450

\bibitem[{{Li} {et~al.}(2001{\natexlab{b}}){Li}, {Filippenko}, {Treffers},
  {Riess}, {Hu}, \& {Qiu}}]{li01b}
{Li} W., {Filippenko} A.~V., {Treffers} R.~R., {Riess} A.~G., {Hu} J., {Qiu}
  Y., 2001{\natexlab{b}}, \apj, 546, 734

\bibitem[{{Li} {et~al.}(2001{\natexlab{a}})}]{li01:00cx}
{Li} W. {et~al.}, 2001{\natexlab{a}}, \pasp, 113, 1178

\bibitem[{{Li} {et~al.}(2003)}]{li03}
{Li} W. {et~al.}, 2003, \pasp, 115, 453

\bibitem[{{Li} {et~al.}(2000)}]{li00}
{Li} W.~D. {et~al.}, 2000, in
Cosmic Explosions, ed. S.~S. {Holt}, W.~W. {Zhang} 
 (New York: AIP), 103

\bibitem[{{Li} {et~al.}(2011{\natexlab{a}})}]{li11a}
{Li} W., {et~al.}  2011{\natexlab{a}}, \mnras, 412, 1441 

\bibitem[{{Li} {et~al.}(2011{\natexlab{b}})}]{li11b}
{Li} W., {Chornock} R., {Leaman} J.,{ Filippenko} A.~V.,{ Poznanski} D., {Wang} X., {Ganeshalingam} M., {Mannucci} F.,
 2011{\natexlab{b}}, \mnras, 412, 1473

\bibitem[{{Lira} {et~al.}(1998)}]{lira98}
{Lira} P. {et~al.}, 1998, \aj, 115, 234

\bibitem[{{Livio}(2000)}]{livio00}
{Livio} M., 2000, in Type Ia Supernovae:
   Observations and Theory, ed. J. C. Niemeyer, J. W. Truran (Cambridge:
   Cambridge Univ. Press), 33

\bibitem[{{Maeda} {et~al.}(2010)}]{maeda10}
{Maeda} K. {et~al.}, 2010, \nat, 466, 82

\bibitem[{{Maeda} {et~al.}(2011){Maeda} {et~al.}(2011)}]{maeda11}
{Maeda} K. {et~al.},  2011, \mnras, 413, 3075 (M11)

\bibitem[{{Matzner} \& {McKee}(1999)}]{matzner99}
{Matzner} C.~D., {McKee} C.~F., 1999, \apj, 510, 379

\bibitem[{{Mazzali} {et~al.}(2005){Mazzali}, {Benetti}, {Stehle}, {Branch},
  {Deng}, {Maeda}, {Nomoto}, \& {Hamuy}}]{mazzali05}
{Mazzali} P.~A., {Benetti} S., {Stehle} M., {Branch} D., {Deng} J., {Maeda} K.,
  {Nomoto} K., {Hamuy} M., 2005, \mnras, 357, 200

\bibitem[{{Miller} \& {Stone}(1993)}]{miller93}
{Miller} J.~S., {Stone} R.~P.~S., 1993, {Lick Obs. Tech. Rep. 66} (Santa Cruz:
  Lick Obs.)

\bibitem[{{Modjaz} {et~al.}(2009){Modjaz}, {Li}, {Butler}, {Chornock},
  {Perley}, {Blondin}, {Bloom}, {Filippenko}, {Kirshner}, {Kocevski},
  {Poznanski}, {Hicken}, {Foley}, {Stringfellow}, {Berlind}, {Barrado y
  Navascues}, {Blake}, {Bouy}, {Brown}, {Challis}, {Chen}, {de Vries},
  {Dufour}, {Falco}, {Friedman}, {Ganeshalingam}, {Garnavich}, {Holden},
  {Illingworth}, {Lee}, {Liebert}, {Marion}, {Olivier}, {Prochaska},
  {Silverman}, {Smith}, {Starr}, {Steele}, {Stockton}, {Williams}, \&
  {Wood-Vasey}}]{modjaz09}
{Modjaz} M. {et al.}, 2009, \apj, 702, 226

\bibitem[{{Modjaz} {et~al.}(2001){Modjaz}, {Li}, {Filippenko}, {King},
  {Leonard}, {Matheson}, {Treffers}, \& {Riess}}]{modjaz01:98de}
{Modjaz} M., {Li} W., {Filippenko} A.~V., {King} J.~Y., {Leonard} D.~C.,
  {Matheson} T., {Treffers} R.~R., {Riess} A.~G., 2001, \pasp, 113, 308

\bibitem[{{Pastorello} {et~al.}(2007)}]{pastorello07}
{Pastorello} A. {et~al.}, 2007, \mnras, 376, 1301

\bibitem[{{Patat} {et~al.}(1996){Patat}, {Benetti}, {Cappellaro}, {Danziger},
  {Della Valle}, {Mazzali}, \& {Turatto}}]{patat96}
{Patat} F., {Benetti} S., {Cappellaro} E., {Danziger} I.~J., {Della Valle} M.,
  {Mazzali} P.~A., {Turatto} M., 1996, \mnras, 278, 111

\bibitem[{{Perlmutter} {et~al.}(1999)}]{perlmutter99}
{Perlmutter} S. {et~al.}, 1999, \apj, 517, 565

\bibitem[{{Phillips}(1993)}]{phillips93}
{Phillips} M.~M., 1993, \apjl, 413, L105

\bibitem[{{Phillips} {et~al.}(2006)}]{phillips06}
{Phillips} M.~M. {et~al.}, 2006, \aj, 131, 2615

\bibitem[{{Phillips} {et~al.}(2007)}]{phillips07}
{Phillips} M.~M. {et~al.}, 2007, \pasp, 119, 360

\bibitem[{{Pignata} {et~al.}(2008)}]{pignata08}
{Pignata} G. {et~al.}, 2008, \mnras, 388, 971

\bibitem[{{Pskovskii}(1984)}]{pskovskii84}
{Pskovskii} Yu.~P., 1984, \sovast, 28, 658

\bibitem[{{Riess} {et~al.}(1998)}]{riess98}
{Riess} A.~G. {et~al.}, 1998, \aj, 116, 1009

\bibitem[{{Riess} {et~al.}(1999{\natexlab{a}})}]{riess99a}
{Riess} A.~G. {et~al.}, 1999{\natexlab{a}}, \aj, 117, 707

\bibitem[{{Riess} {et~al.}(1999{\natexlab{b}})}]{riess99b}
{Riess} A.~G. {et~al.}, 1999{\natexlab{b}}, \aj, 118, 2675 (R99)

\bibitem[{{Riess} {et~al.}(2005)}]{riess05}
{Riess} A.~G. {et~al.}, 2005, \apj, 627, 579

\bibitem[{{Riess} {et~al.}(2007)}]{riess07}
{Riess} A.~G. {et~al.}, 2007, \apj, 659, 98

\bibitem[{{Schlegel} {et~al.}(1998){Schlegel}, {Finkbeiner}, \&
  {Davis}}]{schlegel98}
{Schlegel} D.~J., {Finkbeiner} D.~P., {Davis} M., 1998, \apj, 500, 525

\bibitem[{{Shen} \& {Bildsten}(2009)}]{shen09}
{Shen} K.~J., {Bildsten} L., 2009, \apj, 699, 1365

\bibitem[{{Silverman} {et~al.}(2010){Silverman}, {Ganeshalingam}, {Li},
  {Filippenko}, {Miller}, \& {Poznanski}}]{silverman10}
{Silverman} J.~M., {Ganeshalingam} M., {Li} W., {Filippenko} A.~V., {Miller}
  A.~A., {Poznanski} D., 2010, \mnras,  410, 585

\bibitem[{{Soderberg} {et~al.}(2008){Soderberg}, {Berger}, {Page}, {Schady},
  {Parrent}, {Pooley}, {Wang}, {Ofek}, {Cucchiara}, {Rau}, {Waxman}, {Simon},
  {Bock}, {Milne}, {Page}, {Barentine}, {Barthelmy}, {Beardmore}, {Bietenholz},
  {Brown}, {Burrows}, {Burrows}, {Byrngelson}, {Cenko}, {Chandra}, {Cummings},
  {Fox}, {Gal-Yam}, {Gehrels}, {Immler}, {Kasliwal}, {Kong}, {Krimm},
  {Kulkarni}, {Maccarone}, {M{\'e}sz{\'a}ros}, {Nakar}, {O'Brien}, {Overzier},
  {de Pasquale}, {Racusin}, {Rea}, \& {York}}]{soderberg08}
{Soderberg} A.~M. {et al.}, 2008, \nat, 453, 469

\bibitem[{{Stanishev} {et~al.}(2007)}]{stanishev07}
{Stanishev} V. {et~al.}, 2007, \aap, 469, 645

\bibitem[{{Stritzinger} {et~al.}(2002)}]{stritzinger02}
{Stritzinger} M. {et~al.}, 2002, \aj, 124, 2100

\bibitem[{{Strolger} {et~al.}(2002){Strolger}, {Smith}, {Suntzeff}, {Phillips},
  {Aldering}, {Nugent}, {Knop}, {Perlmutter}, {Schommer}, {Ho}, {Hamuy},
  {Krisciunas}, {Germany}, {Covarrubias}, {Candia}, {Athey}, {Blanc},
  {Bonacic}, {Bowers}, {Conley}, {Dahl{\'e}n}, {Freedman}, {Galaz}, {Gates},
  {Goldhaber}, {Goobar}, {Groom}, {Hook}, {Marzke}, {Mateo}, {McCarthy},
  {M{\'e}ndez}, {Muena}, {Persson}, {Quimby}, {Roth}, {Ruiz-Lapuente},
  {Seguel}, {Szentgyorgyi}, {von Braun}, {Wood-Vasey}, \& {York}}]{strolger02}
{Strolger} L. {et al.}, 2002, \aj, 124, 2905

\bibitem[{{Strovink}(2007)}]{strovink07}
{Strovink} M., 2007, \apj, 671, 1084 (S07)

\bibitem[{{Wang} {et~al.}(2006){Wang}, {Baade}, {H{\"o}flich}, {Wheeler},
  {Kawabata}, {Khokhlov}, {Nomoto}, \& {Patat}}]{wang06}
{Wang} L., {Baade} D., {H{\"o}flich} P., {Wheeler} J.~C., {Kawabata} K.,
  {Khokhlov} A., {Nomoto} K., {Patat} F., 2006, \apj, 653, 490

\bibitem[{{Wang} {et~al.}(2009)}]{wang09}
{Wang} X. {et~al.}, 2009, \apjl, 699, L139

\bibitem[{{Webbink}(1984)}]{webbink84}
{Webbink} R.~F., 1984, \apj, 277, 355

\bibitem[{{Whelan} \& {Iben}(1973)}]{whelan73}
{Whelan} J., {Iben} I., Jr., 1973, \apj, 186, 1007

\bibitem[{{Wood-Vasey} {et~al.}(2007)}]{wood-vasey07}
{Wood-Vasey} W.~M. {et~al.}, 2007, \apj, 666, 694

\bibitem[{{Wood-Vasey} {et~al.}(2008)}]{wood-vasey08}
{Wood-Vasey} W.~M. {et~al.}, 2008, \apj, 689, 377

\bibitem[{{Yamanaka} {et~al.}(2009)}]{yamanaka09}
{Yamanaka} M. {et~al.}, 2009, \apjl, 707, L118

\bibitem[{{Zhang} {et~al.}(2010)}]{zhang10}
{Zhang} T. {et~al.}, 2010, \pasp, 122, 1

\end{thebibliography}
\end{document}